\documentclass{article}
\usepackage{amsmath}

\input{tcilatex}

\begin{document}

\begin{center}
\textbf{Electric and magnetic fields as explicitly observer dependent }

\textbf{four-dimensional vectors and their Lorentz transformations }

\textbf{according to Minkowski - Ivezi\'{c}\bigskip \medskip }

Tomislav Ivezi\'{c}$\bigskip $

\textit{Ru%
\mbox
 {\it{d}\hspace{-.15em}\rule[1.25ex]{.2em}{.04ex}\hspace{-.05em}}er Bo\v{s}%
kovi\'{c} Institute, P.O.B. 180, 10002 Zagreb, Croatia}

E-mail: \textit{ivezic@irb.hr}\bigskip \bigskip
\end{center}

In this paper a geometric approach to the special relativity (SR) is used
that is called the \textquotedblleft invariant special
relativity\textquotedblright\ (ISR). In the ISR it is considered that in the
four-dimensional (4D) spacetime physical laws are geometric, coordinate-free
relationships between the 4D geometric, coordinate-free quantities. It is
mathematicaly proved that in the ISR the electric and magnetic fields are
properly defined vectors on the 4D spacetime. According to the first proof
the dimension of a vector field is mathematicaly determined by the dimension
of its domain. Since the electric and magnetic fields are defined on the 4D
spacetime they are properly defined 4D vectors, the 4D geometric quantities
(GQs). As shown in an axiomatic geometric formulation of electromagnetism
with only one axiom, the field equation for the bivector field $F$ [33], [T.
Ivezi\'{c}, Found. Phys. Lett. \textbf{18,} 401 (2005), arXiv:
physics/0412167], the primary quantity for the whole electromagnetism is the
bivector field $F$. The electric and magnetic fields 4D vectors $E$ and $B$
are determined in a mathematically correct way in terms of $F$\ and the 4D
velocity vector $v$\ of the observer who measures $E$ and $B$ fields.
Furthermore, the proofs are presented that under the mathematicaly correct
Lorentz transformations, which are first derived by Minkowski and reinvented
and generalized in terms of 4D GQs, e.g., in [23], [T. Ivezi\'{c}, Phys.
Scr. \textbf{82,} 055007 (2010)], the electric field 4D vector transforms as
any other 4D vector transforms, i.e., again to the electric field 4D vector;
there is no mixing with the magnetic field 4D vector $B$, as in the usual
transformations (UT) of the 3D fields. Different derivations of these UT of
the 3D fields are discussed and objected from the ISR viewpoint. This
formulation with the 4D GQs is in a true agreement, independent of the
chosen inertial reference frame and of the chosen system of coordinates in
it, with experiments in electromagnetism, e.g., the motional emf. It is
shown that the theory with the 4D fields is always in agreement with the
principle of relativity, whereas it is not the case with the usual approach
with the 3D quantities and their UT.\bigskip

\noindent PACS numbers: 03.30.+p, 03.50.De\bigskip \bigskip

\noindent \textbf{1.} \textbf{Introduction}\bigskip

\noindent Both, in the prerelativistic physics and in Einstein's formulation
of special relativity (SR) [1] it is considered that the electric and
magnetic fields are the three-dimensional (3D) vectors $\mathbf{E(r,}t%
\mathbf{)}$ and $\mathbf{B(r,}t\mathbf{)}$. In the whole physical literature
after [1] the usual transformations (UT) of the 3D vectors $\mathbf{E}$ and $%
\mathbf{B}$,\ the last equations in \S 6., II. Electrodynamical Part, [1]
or, e.g., Eqs. (11.148) and (11.149) in [2], are always considered to be the
relativistically correct Lorentz transformations (LT) (boosts) of $\mathbf{E}
$ and $\mathbf{B}$. Here, in the whole paper, under the name LT we shall
only consider boosts. They are first derived by Lorentz [3] and Poincar\'{e}
[4] (see also two fundamental Poincar\'{e}'s papers with notes by Logunov
[5]) and independently by Einstein [1]. Then, they are subsequently derived
and quoted in almost every textbook and paper on relativistic
electrodynamics. According to these UT, the transformed 3D vector $\mathbf{E}%
^{\prime }$ is expressed by the mixture of the 3D vectors $\mathbf{E}$ and $%
\mathbf{B}$, Eq. (11.149) in [2], i.e., Eq. (\ref{JCB}) here

\begin{eqnarray}
\mathbf{E}^{\prime } &=&\gamma (\mathbf{E}+\mathbf{\beta \times }c\mathbf{B)-%
}(\gamma ^{2}/(1+\gamma ))\mathbf{\beta (\beta \cdot E),}  \notag \\
\mathbf{B}^{\prime } &=&\gamma (\mathbf{B}-(1/c)\mathbf{\beta \times E)-}%
(\gamma ^{2}/(1+\gamma ))\mathbf{\beta (\beta \cdot B),}  \label{JCB}
\end{eqnarray}%
where $\mathbf{E}^{\prime }$, $\mathbf{E}$, $\mathbf{\beta }$ and $\mathbf{B}%
^{\prime }$, $\mathbf{B}$ are all 3D vectors. It is visible from (\ref{JCB})
that, e.g., the electric field $\mathbf{E}$ in one frame is
\textquotedblleft seen\textquotedblright\ as slightly changed electric field 
$\mathbf{E}^{\prime }$ and an \emph{induced magnetic field} $\mathbf{B}%
^{\prime }$ in a relatively moving inertial frame. Henceforward, these UT of
the 3D vectors $\mathbf{E}$ and $\mathbf{B}$ (\ref{JCB}) will be called the
Lorentz-Poincar\'{e}-Einstein transformations (LPET), according to
physicists who discovered them.

In this paper we shall deal with a geometric approach to the SR, which is
called the invariant special relativity\ (ISR).\emph{\ In the ISR it is
considered that in the 4D spacetime physical laws are geometric,
coordinate-free relationships between the 4D geometric, coordinate-free
quantities.}

These 4D geometric quantities (GQs) are well-defined both theoretically and 
\emph{experimentally;} they have an independent physical reality. The
principle of relativity is automatically satisfied if the physical laws are
expressed in terms of 4D GQs. It is not so in the SR [1] in which \emph{the
principle of relativity is postulated }outside the mathematical formulation
of the theory and it is supposed that it holds for physical laws expressed
in terms of 3D quantities. In the ISR physical quantities are represented by
the abstract, coordinate-free, 4D GQs. The coordinate-free 4D GQs will be
called the abstract quantities (AQs). If some basis in 4D spacetime has been
introduced, these AQs are represented as 4D coordinate-based geometric
quantities (CBGQs) comprising both components and \emph{a basis}. Every 4D
CBGQ is invariant under the passive LT; the components transform by the LT
and the basis by the inverse LT leaving the whole CBGQ unchanged. This is
the reason for the name ISR. The invariance of a 4D CBGQ under the passive
LT reflects the fact that such mathematical, invariant, 4D GQ represents 
\emph{the same physical quantity} for relatively moving inertial observers
and for different bases chosen by them. From the mathematical viewpoint
there is no need to introduce CBGQs. However, physicists cannot measure AQs
and therefore it is necessary to introduce CBGQs in order to be able to
compare the results of experiments with the theory.

In contrast to the ISR the usual SR [1] deals with the Lorentz contraction,
the time dilation and the LPET of the 3D vectors $\mathbf{E}$ and $\mathbf{B}
$, (\ref{JCB}). However, as shown in [6 - 10] and Appendix here, e.g., the
Lorentz contraction is ill-defined in the 4D spacetime; it is
synchronization dependent and consequently it is not an intrinsic
relativistic effect. (Observe that in the second paper in [8], the incorrect
quadrupole field of the stationary current loop from the published version
is replaced by the dipole field. Therefore, henceforward, if referred to [8]
I mean that the corrected version has to be taken into account.) The LT have
nothing in common with the Lorentz contraction; the LT cannot connect two 
\emph{spatial lengths} that are simultaneously determined for relatively
moving inertial observers. In the SR the spatial length is defined \emph{as
the spatial distance between two spatial points on the (moving) object
measured by simultaneity in the rest frame of the observer.} \emph{The rest
length and the Lorentz contracted length are not the same 4D quantity for
relatively moving observers and they are not related by the LT, since the
transformed length} $L_{0}(1-\beta ^{2})^{1/2}$ \emph{is different set of
events in the 4D spacetime than the rest length} $L_{0}$, see in [6] Fig. 3.
for the Lorentz contraction and Fig. 4. for the time dilation. Rohrlich [11]
named the Lorentz contraction and other transformations which do not refer
to the same quantity as the \textquotedblleft apparent\textquotedblright\
transformations (AT), whereas the transformations which refer to the same
quantity as the \textquotedblleft true\textquotedblright\ transformations,
e.g., the LT. Hence, the other name for the ISR is the \textquotedblleft
True transformations relativity\textquotedblright\ (\textquotedblleft TT
relativity\textquotedblright ), which is used, e.g., in [6 - 9]. In the 4D
spacetime, as shown in detail in [6 - 10], instead of the Lorentz
contraction and the time dilation one has to consider the 4D GQs, the
position vectors $x_{A}$, $x_{B},$ of the events $A$ and $B$, respectively,
the distance vector $l_{AB}=x_{B}-x_{A}$ and the spacetime length, which all
properly transform under the LT. The essential feature of the geometric
approach is that \emph{any abstract 4D geometric quantity} \emph{(or a 4D
CBGQ)}, e.g., the distance vector $l_{AB}$ \emph{is} \emph{only one
quantity, the same quantity in the 4D spacetime} for all relatively moving
frames of reference and for all systems of coordinates that are chosen in
them, see in [6], Fig. 1. for the spacetime length for a moving rod and Fig.
2. for the spacetime length for a moving clock. In [7] it is explicitly
shown that all well-known experiments that test special relativity, e.g.,
the \textquotedblleft muon\textquotedblright\ experiment, the
Michelson-Morley type experiments, the Kennedy-Thorndike type experiments
and the Ives-Stilwell type experiments are in a complete agreement,
independently of the chosen synchronization, with the 4D geometric approach,
whereas it is not the case with Einstein's approach [1] with the Lorentz
contraction and the time dilation if the \textquotedblleft
radio\textquotedblright\ (\textquotedblleft r\textquotedblright )
synchronization is used; see two papers in arxiv in [7] in which the
\textquotedblleft r\textquotedblright\ synchronization is explicitly used
throughout these two papers. Every synchronization is only a convention and
physics must not depend on conventions, i.e., no experiment should depend on
the chosen synchronization.

Here, in Sec. 2., the geometric algebra formalism [12-15], the standard
basis and the\textbf{\ }$\{r_{\mu }\}$ basis with the \textquotedblleft
r\textquotedblright\ synchronization are briefly discussed. In Secs. 3.1.
and 3.3. it is proved in a mathematically correct way that in the ISR, in
the 4D spacetime, the electric and magnetic fields are properly defined
vectors on the 4D spacetime, the 4D vectors $E$ and $B$. In the whole text $%
E $, $B$ will be simply called -\ vectors - or the 4D vectors, whereas an
incorrect expression, the 3D vector, will still remain for the usual $%
\mathbf{E(r,}t\mathbf{)}$, $\mathbf{B(r,}t\mathbf{)}$, e.g., from Eq. (\ref%
{JCB}). Mathematically, the 3D $\mathbf{E(r,}t\mathbf{)}$, $\mathbf{B(r,}t%
\mathbf{)}$ are not properly defined vectors on the 4D spacetime.\ In Secs.
4.1. and 4.2. it is proved that from the ISR viewpoint the UT of the 3D
fields, i.e., the LPET, are not the mathematically correct LT, because the
LT are properly defined on the 4D spacetime and cannot transform the 3D
quantities. The LT transform the electric field vector in the same way as
any other vector transforms, i.e., again to the electric field vector.
Minkowski, in Sec. 11.6 in [16], was the first who introduced the electric
and magnetic fields as 4D vectors and derived their correct LT but only with
components implicitly taken in the standard basis. This is reinvented and
generalized in terms of 4D GQs in [17] - [23], see also the discussion in
[10]. Here, the LT of the 4D vectors $E$ and $B$ will be called the
Minkowski-Ivezi\'{c} LT (MILT). Particularly, in [23], the comparison of our
approach with 4D GQs and Minkowski's results is presented in detail. Note,
however, that Minkowski never explicitly wrote the LT of the 4D $E$ and $B$,
Eqs. (\ref{aec}) - (\ref{el}) here and he never applied these
transformations. Sections 3.1., 3.3., 4.1. and 4.2. are the central sections
and they contain the most important results that are obtained in this paper.
In Secs. 5.1. and 5.2., for the reader's convenience, the derivations of the
UT of the 3D $\mathbf{E}$ and $\mathbf{B}$, the LPET, and the LT of the 4D $%
E $ and $B$, the MILT, are compared using matrices. In Sec. 7., we discuss
the derivation of the LPET from the usual covariant approaches, e.g., from
[2]. In Sec. 8., the derivation of the LPET from the textbook by Blandford
and Thorne (BT) [24] is discussed and objected from the ISR viewpoint. In
[24], in contrast to, e.g., [2], [25], a geometric viewpoint is adopted; 
\emph{the physical laws are stated as geometric, coordinate-free
relationships between the geometric, coordinate-free quantities. }%
Particularly, in Sec. 1.10 in [24], \emph{it is discussed the nature of
electric and magnetic fields and they are considered to be the 4D fields.}
But, nevertheless, BT also derived the UT of the 3D vectors $\mathbf{E}$ and 
$\mathbf{B}$, the LPET, their Eq. (1.113), and not the correct LT of the 4D
fields, the MILT, Eqs. (\ref{aec}) - (\ref{el}) here. They have \emph{not }%
noticed that under the LT the electric field 4D vector must transform as any
other 4D vector transforms. In Sec. 10., the electromagnetic field of a
point charge in uniform motion is investigated and it is explicitly shown
that 1) the primary quantity is the bivector $F$\ (Eqs. (\ref{fcf}) and (\ref%
{fmn})) and 2) that the observer dependent 4D vectors $E$ and $B$, Eq. (\ref%
{ecb}), correctly describe both the electric and magnetic fields for all
relatively moving inertial observers and for all bases chosen by them. In
Sec. 11., a brief discussion is presented of the comparison with the
experiments on the motional emf. It is shown that the theory with the 4D GQs
and their MILT, Eqs. (\ref{aec}) - (\ref{el}) here, is always in agreement
with the principle of relativity, whereas it is not the case with the usual
approach with the 3D quantities and their UT, the LPET. In Sec. 12., the
discussion of the obtained results is presented and the conclusions are
given.\bigskip \bigskip

\noindent \textbf{2. The geometric algebra formalism - the }$\{r_{\mu }\}$ 
\textbf{basis with the \textquotedblleft r\textquotedblright\ synchronization%
}\bigskip

\noindent Here, as already said, we shall deal either with the abstract,
coordinate-free 4D GQs, AQs, or with their representations in some basis, 4D
CBGQs comprising \emph{both} components and \emph{a basis}, e.g., the
position vector, $x=x^{\nu }\gamma _{\nu }$. We shall use the geometric
algebra formalism, see, e.g., [12-15]. The geometric (Clifford) product of
two multivectors $A$ and $B$ is written by simply juxtaposing multivectors $%
AB$. For vectors $a$ and $b$ the geometric product $ab$\ decomposes as $%
ab=a\cdot b+a\wedge b$, where the inner product $a\cdot b$\ is $a\cdot
b\equiv (1/2)(ab+ba)$ and the outer (or exterior) product $a\wedge b\ $is\ $%
a\wedge b\equiv (1/2)(ab-ba)$.\ For the reader's convenience, all equations
will be also written with CBGQs in the standard basis. Therefore, the
knowledge of the geometric algebra is not required for the understanding of
this presentation. The standard basis $\left\{ \gamma _{\mu }\right\} $ is a
right-handed orthonormal frame of vectors in the Minkowski spacetime $M^{4}$
with $\gamma _{0}$ in the forward light cone, $\gamma _{0}^{2}=1$ and $%
\gamma _{k}^{2}=-1$ ($k=1,2,3$). The $\gamma _{\mu }$ generate by
multiplication a complete basis for the spacetime algebra: $1$, $\gamma
_{\mu }$, $\gamma _{\mu }\wedge \gamma _{\nu }$, $\gamma _{\mu }\gamma _{5}$%
, $\gamma _{5}$ ($2^{4}=16$ independent elements). $\gamma _{5}$ is the
right-handed unit pseudoscalar, $\gamma _{5}=\gamma _{0}\wedge \gamma
_{1}\wedge \gamma _{2}\wedge \gamma _{3}$. Any multivector can be expressed
as a linear combination of these 16 basis elements of the spacetime algebra.
The $\left\{ \gamma _{\mu }\right\} $ basis corresponds to Einstein's system
of coordinates in which the Einstein synchronization of distant clocks [1]
and Cartesian space coordinates $x^{i}$ are used in the chosen inertial
frame of reference (IFR). Here, as in [6, 7, 8, 26], we shall also introduce
another basis, the $\{r_{\mu }\}$ basis with the \textquotedblright
everyday\textquotedblright\ or \textquotedblright radio\textquotedblright\
(\textquotedblright r\textquotedblright ) synchronization [27]. The
\textquotedblleft r\textquotedblright\ synchronization is commonly used in
everyday life, [27].

The \textquotedblright r\textquotedblright\ synchronization differs from the
Einstein synchronization by the different procedure for the synchronization
of distant clocks. Different synchronizations are determined by the
parameter $\varepsilon $ in the relation $t_{2}=t_{1}+\varepsilon
(t_{3}-t_{1})$, where $t_{1}$ and $t_{3}$ are the times of departure and
arrival, respectively, of the light signal, read by the clock at $A$, and $%
t_{2}$ is the time of reflection at $B$, read by the clock at $B$, that has
to be synchronized with the clock at $A$. In Einstein's synchronization
convention $\varepsilon =1/2.$ In the \textquotedblright
r\textquotedblright\ synchronization $\varepsilon =0$ and thus, in contrast
to the Einstein synchronization, there is an absolute simultaneity. As
explained in [27]: \textquotedblright For if we turn on the radio and set
our clock by the standard announcement \textquotedblright ...at the sound of
the last tone, it will be 12 o'clock\textquotedblright , then we have
synchronized our clock with the studio clock in a manner that corresponds to
taking $\varepsilon =0$ in $t_{2}=t_{1}+\varepsilon (t_{3}-t_{1}).$%
\textquotedblright\ The \textquotedblright r\textquotedblright\
synchronization is an assymetric synchronization which leads to an assymetry
in the coordinate, one-way, speed of light. However from the physical point
of view the \textquotedblright r\textquotedblright\ synchronization is
completely equivalent to the Einstein synchronization. This also holds for
all other permissible synchronizations. Such situation really happens in the
ISR\ since the ISR\ deals with the coordinate-free 4D GQs or with the 4D
CBGQs. Thus the ISR\ deals on the same footing with all possible systems of
coordinates of the chosen IFR. As a consequence \emph{the second Einstein
postulate referred to the constancy of the coordinate velocity of light, in
general, does not hold in the ISR}$.$ Namely, only with Einstein's
synchronization the coordinate, one-way, speed of light is isotropic and
constant.

The basis vectors in the $\{\gamma _{\mu }\}$ basis and the $\{r_{\mu }\}$
basis are constructed as in [27] and [8]. The temporal basis vector $\gamma
_{0}$ is the unit vector directed along the world line of the clock at the
origin. The spatial basis vectors by definition connect \emph{simultaneous}
events, the event \textquotedblright clock at rest at the origin reads 0
time\textquotedblright\ with the event \textquotedblright clock at rest at
unit distance from the origin reads 0 time,\textquotedblright\ and thus they
are synchronization-dependent. The spatial basis vector $\gamma _{i}$
connects two above mentioned simultaneous events when Einstein's
synchronization ($\varepsilon =1/2$) of distant clocks is used. The temporal
basis vector $r_{0}$ is the same as $\gamma _{0}.$ The spatial basis vector $%
r_{i}$ connects two above mentioned simultaneous events when
\textquotedblright radio\textquotedblright\ clock synchronization ($%
\varepsilon =0$) of distant clocks is used. The spatial basis vectors, e.g., 
$r_{1},r_{1}^{\prime },r_{1}^{\prime \prime }..$ for relatively moving IFRs
are parallel and directed along an (observer-independent) light line. Hence,
two events that are everyday (\textquotedblright r\textquotedblright )
simultaneous in some IFR $S$ are also \textquotedblright
r\textquotedblright\ simultaneous for all other relatively moving IFRs.

It is shown in a simple way in [27], see Eq. (3) and Figure 2, that the unit
vectors in the $\{\gamma _{\mu }\}$ basis and the $\{r_{\mu }\}$ basis are
connected as $r_{0}=\gamma _{0}$, $r_{i}=\gamma _{0}+\gamma _{i}$. Hence,
the components $g_{\mu \nu ,r}$ of the metric tensor are $g_{ii,r}=0$, and
all other components are $=1$, see Eq. (4) in [27]. Observe that it is dealt
with 2D spacetime in [27]. Obviously it is completely different than in the $%
\left\{ \gamma _{\mu }\right\} $ basis, i.e. than the Minkowski metric,
which, here, is chosen to be $g_{\mu \nu }=diag(1,-1,-1,-1)$. Note that in
[6] - [9] the Minkowski metric is $g_{\mu \nu }=diag(-1,1,1,1)$. Then,
according to Eq. (4) from [6], one can use $g_{\mu \nu ,r}$ to find the
transformation matrix $R_{\;\nu }^{\mu }$ that connects the components in
the $\left\{ \gamma _{\mu }\right\} $ and the $\{r_{\mu }\}$ bases. The only
components that are different from zero are%
\begin{equation}
R_{\;\mu }^{\mu }=-R_{\;i}^{0}=1.  \label{er}
\end{equation}%
The inverse matrix $(R_{\;\nu }^{\mu })^{-1}$ connects the \textquotedblleft
old\textquotedblright\ basis, $\left\{ \gamma _{\mu }\right\} $, with the
\textquotedblleft new\textquotedblright\ one, $\{r_{\mu }\}$. In [6], $%
R_{\;\nu }^{\mu }$ is obtained from Logunov's expression for the
transformation matrix $\lambda _{\nu }^{\overline{\mu }}$ connecting (in his
interpretation) a physicaly measurable tensor with the coordinate one (A.A.
Logunov, \textit{Lectures in the theory of relativity and gravity. A
present-day analysis of the problem }(Nauka, Moskva, 1987) (in Russian).) In
the mentioned approach there are physical and coordinate quantities in the
considered IFR. However, in our interpretation, his \textquotedblright
physicaly measurable tensor\textquotedblright\ corresponds to the tensor
written in the Einstein basis $\left\{ \gamma _{\mu }\right\} $ of a given
IFR, and the coordinate one corresponds to some arbitrary basis of the same
IFR. Hence, his matrix $\lambda _{\nu }^{\overline{\mu }}$ can be
interpreted as the transformation matrix between some arbitrary basis and
the Einstein\ basis. The elements of $\lambda _{\nu }^{\overline{\mu }}$,
which are different from zero, are $\lambda _{0}^{\overline{0}%
}=(g_{00})^{1/2},$ $\lambda _{i}^{\overline{0}}=(g_{0i})(g_{00})^{-1/2},%
\quad \lambda _{i}^{\overline{i}}=\left[ g_{ii}+(g_{0i})^{2}/g_{00}\right]
^{1/2}$. We actually need the inverse transformation $(\lambda _{\nu }^{%
\overline{\mu }})^{-1}$ that will be denoted $T_{\;\nu }^{\mu }$ as in [6].
Then the elements different from zero of the matrix $T_{\;\nu }^{\mu }$ are
determined by the basis components $g_{\mu \nu }$ of the metric tensor in
that arbitrary basis, e.g., the $\{r_{\mu }\}$ basis,%
\begin{eqnarray}
T_{\;0}^{0} &=&(g_{00})^{-1/2},\quad T_{\;i}^{0}=(g_{0i})(-g_{00})^{-1}\left[
g_{ii}+(g_{0i})^{2}/g_{00}\right] ^{-1/2}  \notag \\
T_{\;i}^{i} &=&\left[ g_{ii}+(g_{0i})^{2}/g_{00}\right] ^{-1/2}  \label{lamb}
\end{eqnarray}%
The transformation matrix $R_{\;\nu }^{\mu }$\ , i.e., $T_{\;\nu ,r}^{\mu }$
is then easily obtained from $T_{\;\nu }^{\mu }$ (\ref{lamb}) and the known $%
g_{\mu \nu ,r}$. We note that in SR, i.e., in the theory of the flat
spacetime, any specific $g_{\mu \nu }$ (for the specific basis) can be
transformed to the Minkowski metric. It can be accomplished by means of the
matrix $(T_{\;\nu }^{\mu })^{-1}$; for example, $g_{\mu \nu ,r}$ is
transformed by the matrix $(T_{\;\nu ,r}^{\mu })^{-1}$ to the Minkowski $%
g_{\mu \nu }$.

In the $\{r_{\mu }\}$ basis the components of any vector are connected in
the same way as the components of the position vector $x$ are connected,
i.e., as $x_{r}^{\mu }$=$R_{\;\nu }^{\mu }x^{\nu }$,%
\begin{equation}
x_{r}^{0}=x^{0}-x^{1}-x^{2}-x^{3},\quad x_{r}^{i}=x^{i}.  \label{ptr}
\end{equation}%
This reveals that \emph{in the} $\{r_{\mu }\}$\emph{\ basis the space} $%
\mathbf{r}$ \emph{and the time} $t$\ \emph{cannot be separated}; the
\textquotedblleft 3+1 split\textquotedblright\ of the spacetime into space +
time is \emph{impossible}. Note that there is the zeroth component of $x$\
in the $\{r_{\mu }\}$ basis, $x_{r}^{0}\neq 0$, even if in the standard
basis $x^{0}=0$, but the spatial components $x^{i}\neq 0$. This means that 
\emph{in the 4D spacetime only the position vector} $x$, $x=x^{\mu }\gamma
_{\mu }=x_{r}^{\mu }r_{\mu }$, \emph{is properly defined quantity}. In
general, the position in the 3D space $\mathbf{r}$ and the time $t$\ have
not an independent reality in the 4D spacetime. Although the Einstein and
the \textquotedblleft r\textquotedblright\ synchronizations are completely
different they are equally well physical and relativistically correct
synchronizations. \emph{Every synchronization is only a convention and
physics must not depend on conventions.} An important consequence of the
result that in the 4D spacetime $\mathbf{r}$ and $t$\ are not well-defined
is presented in Sec. 4. in [23]. There, it is shown that only the world
parity $W$, $Wx=-x$, is well defined in the 4D spacetime and not the usual $%
T $ and $P$ inversions.

The same result as in Eq. (\ref{ptr}) is also obtained in [27], Eqs. (14a)
and (14b), but in the 2D spacetime. There, it is interpreted as "The
transformation between an observer synchronizing his/her clocks with
Einstein's procedure, and one synchronizing with the `everyday'
procedure."\bigskip \bigskip

\noindent \textbf{3. In the ISR the electric and magnetic fields are well
defined 4D vectors}\bigskip \medskip

\noindent \textit{3.1. Oziewicz's proof}\bigskip

\noindent There is a simple but very strong and completely correct
mathematical argument, which is stated by Oziewicz, e.g., in [28]. There, it
is explained that an individual vector has no dimension; the dimension is
associated with the vector space and with the manifold where this vector is
tangent. Hence, \emph{what is essential for the number of components of a
vector field is the number of variables on which that vector field depends,
i.e., the dimension of its domain.\ In general, the dimension of a vector
field that is defined on a n-dimensional space is equal - n. The electric
and magnetic fields are defined on a 4D space, i.e., the spacetime. They are
always functions of the position vector} $x$. \emph{This means that they are
not the usual 3D fields},\emph{\ }$\mathbf{E(r,}t\mathbf{)}$ \emph{and} $%
\mathbf{B(r,}t\mathbf{)}$, \emph{but they are properly defined vectors on
the 4D spacetime}, $E(x)$ \emph{and} $B(x)$. \emph{This fact determines that
such vector fields, when represented in some basis, have to have four
components (some of them can be zero).} This is a fundamental argument and
it cannot be disputed in any way. It is very surprising that this argument
is not used in physics much earlier. For an exact mathematical proof of that
argument see, e.g., Chapter I in an undergraduate text in mathematics [29]
or Chapter II in an advanced text [30].

The mentioned argument holds in the same measure for the polarization vector 
$P(x)$ and the magnetization vector $M(x)$, which are discussed in detail in
[10, 31, 32]. In [31] the electromagnetic field equations for moving media
are presented, whereas in [32] the constitutive relations and the
magnetoelectric effect for moving media are investigated from the geometric
point of view. In addition to Oziewicz's proof we note that in the 4D
spacetime we \emph{always} have to deal with correctly defined vectors $E(x)$%
, $B(x)$, $P(x)$, $M(x)$, etc. even in the usual static case, i.e., if the
usual 3D fields $\mathbf{E(\mathbf{r})}$, $\mathbf{B(r)}$, ... do not
explicitly depend on the time $t$. The reason is that if \emph{in the 4D
spacetime }the standard basis is used then\emph{\ }the LT cannot transform
the spatial coordinates from one frame only to spatial coordinates in a
relatively moving inertial frame of reference. What is static case for one
inertial observer is not more static case for relatively moving inertial
observer, but a time dependent case. Furthermore, if an observer uses the
\textquotedblleft r\textquotedblright\ synchronization and not Einstein's
synchronization, then, as seen from (\ref{ptr}), the space and time are not
separated and the usual 3D vector $\mathbf{\mathbf{r}}$ is meaningless. If
the principle of relativity has to be satisfied and the physics must be the
same for all inertial observers and for $\{\gamma _{\mu }\}$, $\{r_{\mu }\}$%
, $\{\gamma _{\mu }^{\prime }\}$, etc. bases which they use, then the
properly defined quantity is the position vector $x$, 
\begin{equation}
x=x^{\nu }\gamma _{\nu }=x^{\prime \nu }\gamma _{\nu }^{\prime }=x_{r}^{\nu
}r_{\nu }=x_{r}^{\prime \nu }r_{\nu }^{\prime },  \label{iks}
\end{equation}%
and not $\mathbf{\mathbf{r}}$ and $t$. In (\ref{iks}), the primed quantities
in both bases $\{\gamma _{\mu }\}$ and $\{r_{\mu }\}$ are the Lorentz
transforms of the unprimed ones. For the $\{r_{\mu }\}$ basis and the LT in
that basis see [6]. Consequently, in the 4D spacetime, e.g., the electric
field is properly defined as the vector $E(x)$ for which, in the same way as
in (\ref{iks}), the relation (\ref{erc}) given below holds.\bigskip \medskip

\noindent \textit{3.2. Briefly about the }$F$\textit{\ formulation}\bigskip

\noindent In this section for the sake of completeness and for better
understanding of the whole exposition we briefly repeat main results from
[33]. In [33] an axiomatic geometric formulation of electromagnetism with
only one axiom, the field equation for the bivector field $F$ is
constructed. There, it is shown that the bivector $F=F(x)$, which represent
the electromagnetic field, can be taken as the primary quantity for the
whole electromagnetism. It yields a complete description of the
electromagnetic field and, in fact, there is no need to introduce either the
field vectors or the potentials. If the field equation for $F$\ is written
with AQs it becomes 
\begin{equation}
\partial \cdot F+\partial \wedge F=j/\varepsilon _{0}c,  \label{fef}
\end{equation}%
where the source of the field is the charge-current density vector $j(x)$.
If $j(x)$\ is the sole source of $F$\ then the general solution for $F$\
with AQs\emph{\ }is given by Eq. (8) in [33]. Particularly, in [33], it is
presented the general expression for $F$ for an arbitrary motion of a
charge. It is also specified to the simple case of $F$\ for a point charge
in uniform motion as an AQ, Eq. (\ref{fcf}) below. The components in the
standard basis $F^{\alpha \beta }$ for an arbitrary motion of a charge are
the same as the usual result from Chapter 14 in [2]. Note that in Sec. 2.4.
in [33] the integral form of Eq. (\ref{fef}) is also presented and
discussed. If the equation for $F$\ (\ref{fef}) is written with CBGQs in the 
$\left\{ \gamma _{\mu }\right\} $ basis it becomes 
\begin{equation}
\partial _{\alpha }F^{\alpha \beta }\gamma _{\beta }-\partial _{\alpha }\
^{\ast }F^{\alpha \beta }\gamma _{5}\gamma _{\beta }=(1/\varepsilon
_{0}c)j^{\beta }\gamma _{\beta },  \label{mf}
\end{equation}%
where the usual dual tensor (components) is $^{\ast }F^{\alpha \beta
}=(1/2)\varepsilon ^{\alpha \beta \gamma \delta }F_{\gamma \delta }$. From
that equation one easily finds the usual covariant form (only the basis
components of the 4D GQs in the $\left\{ \gamma _{\mu }\right\} $ basis) of
the field equations as%
\begin{equation}
\partial _{\alpha }F^{a\beta }=j^{\beta }/\varepsilon _{0}c,\quad \partial
_{\alpha }\ ^{\ast }F^{\alpha \beta }=0.  \label{mfc}
\end{equation}%
These two equations for \emph{the components in the standard basis}$\
F^{\alpha \beta }$\ are the equations (11.141) and (11.142) in [2].

In the same paper, [33], it is also shown that this formulation with the $F$%
\ field is in a complete agreement with the Trouton-Noble experiment, i.e.,
in the approach with $F$\ as a 4D GQ there is no Trouton-Noble paradox. It
is clearly visible from [33] and this short presentation that, in principle, 
\emph{the components }$F^{\alpha \beta }$\ of the electromagnetic field
tensor, i.e., of the bivector $F$\ here and in [33], \emph{have nothing to
do with the components of the 3D vectors} $\mathbf{E}$ \emph{and} $\mathbf{B}
$. \emph{The whole} $F$\ \emph{is a physically measurable quantity by the
Lorentz force density}, $K_{(j)}=F\cdot j/c$, \emph{or, for a charge} $q$ 
\emph{by the Lorentz force}

\begin{equation}
K_{L}=(q/c)F\cdot u,  \label{LF}
\end{equation}%
where $u$ is the 4D velocity vector of a charge $q$ (it is defined to be the
tangent to its world line).

It is worth noting that the expression for the Lorentz force density, $%
K_{(j)}=F\cdot j/c$, is directly derived from the field equation for $F$ (%
\ref{fef}). Similarly, in [33], the coordinate-free expressions for the
stress-energy vector $T(n)$ and the quantities derived from $T(n)$, the
energy density $U$ (scalar), the Poynting 4D vector $S$, the momentum
density 4D vector $g=(1/c^{2})S$, the angular momentum density $M$, $%
M=(1/c)T(n)\wedge x$\ (bivector), the local charge conservation law and the
local energy-momentum conservation law are all directly derived from that
field equation (\ref{fef}). In that axiomatic geometric formulation from
[33] $T(n)$\ is the most important quantity for the momentum and energy of
the electromagnetic field. $T(n)$ is a vector-valued linear function on the
tangent space at each spacetime point $x$ describing the flow of
energy-momentum through a hypersurface with normal $n=n(x)$, $%
T(n)=-(\varepsilon _{0}/2)\left[ (F\cdot F)n+2(F\cdot n)\cdot F\right] $. In
Eq. (38) in [33] $T(n)$\ is written in a new form as a sum of $n$-parallel
part ($n-\parallel $) and $n$-orthogonal part ($n-\perp $), $%
T(n)=-(\varepsilon _{0}/2)\left[ (F\cdot F)+2(F\cdot n)^{2}\right]
n-\varepsilon _{0}\left[ (F\cdot n)\cdot F-(F\cdot n)^{2}n\right] $. The
first term is $n-\parallel $ part and it yields the energy density $U$, $%
U=n\cdot T(n)$, $U=-(\varepsilon _{0}/2)\left[ (F\cdot F)+2(F\cdot n)^{2}%
\right] $, whereas the second term is $n-\perp $ part and it is $(1/c)S$,
where $S$\ is the Poynting 4D vector, $S=-\varepsilon _{0}c\left[ (F\cdot
n)\cdot F-(F\cdot n)^{2}n\right] $. and, as can be seen, $n\cdot S=0$. Thus $%
T(n)$ is expressed by $U$ and $S$ as $T(n)=Un+(1/c)S$;

\begin{eqnarray}
T(n) &=&Un+(1/c)S,\quad U=-(\varepsilon _{0}/2)\left[ (F\cdot F)+2(F\cdot
n)^{2}\right] ,  \notag \\
S &=&-\varepsilon _{0}c\left[ (F\cdot n)\cdot F-(F\cdot n)^{2}n\right] .
\label{ust}
\end{eqnarray}

Observe that $T(n)$ as a whole quantity, i.e., the combination of $U$ and $S$
enters into a fundamental physical law, the local energy-momentum
conservation law, $\partial \cdot T(n)=0$, for the free fields. This means,
as stated in [33], that only $T(n)$, as a whole quantity, does have a
physically correct interpretation. In [33] this viewpoint is nicely
illustrated considering \emph{an apparent paradox in the usual 3D formulation%
} in which the 3D Poynting vector $S$ is interpreted as an energy flux due
to the propagation of the 3D fields. If such an interpretation of $S$\ is
adopted then there is a paradox for the case of an uniformly accelerated
charge, e.g., Sec. 6.8 in [2]. In that case, the 3D $S$ is $=0$ (there is no
energy flow) but at the same time the 3D $U$ is $\neq 0$ (there is an energy
density) for the field points on the axis of motion. The obvious question is
how the fields propagate along the axis of motion to give that $U\neq 0$. In
the formulation with 4D GQs the important quantity is $T(n)$ and not $S$ and 
$U$ taken separately. $T(n)$ is $\neq 0$ everywhere on the axis of motion
and \emph{the local energy-momentum conservation law} with $T(n)$ \emph{%
holds everywhere.}\bigskip \medskip

\noindent \textit{3.3. Proof by the use of the decomposition of} $F$\bigskip

\noindent In contrast to the usual covariant approach, which deals with the
identification of components, see in Sec. 7. below, Eqs (\ref{ieb}) and (\ref%
{eb2}), it is possible to construct in a mathematically correct way the 4D
vectors of the electric and magnetic fields using the decomposition of $F$.
There is \emph{a mathematical theorem according to which any antisymmetric
tensor of the second rank can be decomposed into two space-like vectors and
the unit time-like vector.} For the proof of that theorem in geometric terms
see, e.g., [34], [35].

If that theorem is applied to the bivector $F$ then it is obtained that

\begin{equation}
F=E\wedge v/c+(IcB)\cdot v/c,  \label{E2}
\end{equation}%
where the electric and magnetic fields are represented by vectors $E(x)$ and 
$B(x)$, see, e.g., [17 - 19, 33]. The unit pseudoscalar $I$ is defined
algebraically without introducing any reference frame as in [14]. If $I$ is
represented in the $\left\{ \gamma _{\mu }\right\} $ basis it becomes $%
I=\gamma _{0}\wedge \gamma _{1}\wedge \gamma _{2}\wedge \gamma _{3}=\gamma
_{5}$. The vector $v$ in the decomposition (\ref{E2}) is interpreted as the
velocity vector of the observer who measures $E$ and $B$ fields. Then $E(x)$
and $B(x)$ are defined with respect to $v$, i.e., with respect to the
observer, as 
\begin{equation}
E=F\cdot v/c,\quad B=-(1/c)I(F\wedge v/c).  \label{E1}
\end{equation}%
It also holds that $E\cdot v=B\cdot v=0$; both $E$ and $B$ are space-like
vectors. If the decomposition (\ref{E2}) is written with the CBGQs in the $%
\{\gamma _{\mu }\}$ basis it becomes

\begin{equation}
F=(1/2)F^{\mu \nu }\gamma _{\mu }\wedge \gamma _{\nu },\ F^{\mu \nu
}=(1/c)(E^{\mu }v^{\nu }-E^{\nu }v^{\mu })+\varepsilon ^{\mu \nu \alpha
\beta }v_{\alpha }B_{\beta },  \label{fm}
\end{equation}%
where $\gamma _{\mu }\wedge \gamma _{\nu }$ is the bivector basis. If the
equations for $E$ and $B$ (\ref{E1}) are written with the CBGQs in the $%
\{\gamma _{\mu }\}$ basis they become%
\begin{equation}
E=E^{\mu }\gamma _{\mu }=(1/c)F^{\mu \nu }v_{\nu }\gamma _{\mu },\quad
B=B^{\mu }\gamma _{\mu }=(1/2c^{2})\varepsilon ^{\mu \nu \alpha \beta
}F_{\nu \alpha }v_{\beta }\gamma _{\mu }.  \label{ebv}
\end{equation}%
All these relations, (\ref{E2}) - (\ref{ebv}) are the mathematically correct
definitions. They are first reported (only components implicitly taken in
the standard basis) by Minkowski in Sec. 11.6 in [16], see also [23].

Let us introduce the $\gamma _{0}$ - frame; the frame of \textquotedblleft
fiducial\textquotedblright\ observers for which $v=c\gamma _{0}$ and in
which the standard basis is chosen. Therefore, in the $\gamma _{0}$ - frame,
e.g., $E$ becomes $E=F\cdot \gamma _{0}$. It can be shown that in the $%
\gamma _{0}$ - frame $E\cdot \gamma _{0}=B\cdot \gamma _{0}=0$, which means
that $E$ and $B$ are orthogonal to $\gamma _{0}$; they refer to the 3D
subspace orthogonal to the specific timelike direction $\gamma _{0}$. If $E$
and $B$ are written as CBGQs in the standard basis they become%
\begin{eqnarray}
E &=&E^{\mu }\gamma _{\mu }=0\gamma _{0}+F^{i0}\gamma _{i},  \notag \\
B &=&B^{\mu }\gamma _{\mu }=0\gamma _{0}+(1/2c)\varepsilon
^{0ijk}F_{kj}\gamma _{i}.  \label{gnl}
\end{eqnarray}%
Note that $\gamma _{0}=(\gamma _{0})^{\mu }\gamma _{\mu }$ with $(\gamma
_{0})^{\mu }=(1,0,0,0)$. Hence, in the $\gamma _{0}$ - frame the temporal
components of $E$ and $B$ are zero and only the spatial components remain%
\begin{equation}
E^{0}=B^{0}=0,\quad E^{i}=F^{i0},\quad B^{i}=(1/2c)\varepsilon ^{0ijk}F_{kj}.
\label{sko}
\end{equation}%
It is visible from (\ref{gnl}) and (\ref{sko}) that $E^{i}$ and $B^{i}$ are
the same as the components of the 3D $\mathbf{E}$ and $\mathbf{B}$, Sec. 7.,
Eq. (\ref{ieb}) below, i.e., the same as in Eq. (11.137) in [2]. However,
there are very important differences between the identifications (\ref{ieb})
and Eqs. (\ref{gnl}) and (\ref{sko}). The components of $\mathbf{E}$ and $%
\mathbf{B}$ in (\ref{ieb}) are not the spatial components of the 4D
quantities. They transform according to the LPET, Sec. 7., Eq. (\ref{ee})
below. The antisymmetric $\varepsilon $ tensor in (\ref{ieb}) and (\ref{eb2}%
) is a third-rank antisymmetric tensor. On the other hand, the components of 
$E$ and $B$ in (\ref{gnl}) and (\ref{sko}) are the spatial components of the
4D geometric quantities that are taken in the standard basis. They transform
according to the MILT that are given below, Eq. (\ref{LTE}). The
antisymmetric $\varepsilon $ tensor in (\ref{gnl}) and (\ref{sko}) is a
fourth-rank antisymmetric tensor. Furthermore, as shown in Sec. 7., Eqs. (%
\ref{are}) and (\ref{FEr}) below, the identifications (\ref{ieb}) and (\ref%
{eb2}) do not hold in the $\{r_{\mu }\}$ basis. But, the relations (\ref{ebv}%
) hold for any chosen basis, including the $\{r_{\mu }\}$ basis, e.g., 
\begin{equation}
E=E^{\nu }\gamma _{\nu }=E_{r}^{\nu }r_{\nu }=(1/c)F_{r}^{\mu \nu }v_{\nu
,r}r_{\mu }.  \label{efr}
\end{equation}%
This can be easily checked using the above mentioned matrix $R_{\;\nu }^{\mu
}$.\ Thus, for the components of vector $E$ it also holds that 
\begin{equation}
E_{r}^{0}=E^{0}-E^{1}-E^{2}-E^{3},\quad E_{r}^{i}=E^{i}.  \label{ei}
\end{equation}%
From these relations it follows that there is the zeroth component of $E$\
in the $\{r_{\mu }\}$ basis, $E_{r}^{0}\neq 0$, even if it is $=0$ in the
standard basis, $E^{0}=0$, but the spatial components $E^{i}\neq 0$. This
again shows that \emph{the components taken alone are not physical}. The
whole consideration presented here explicitly reveals that in the 4D
spacetime, from the ISR viewpoint, the usual identifications (\ref{ieb}) and
(\ref{eb2}) are not mathematically correct and that in the ISR \emph{the
electric field} $E$ \emph{is a vector (4D} \emph{vector); it is an inner
product of a bivector} $F$\ \emph{and the velocity vector} $v$ \emph{of the
observer who measures fields.}

It is worth mentioning that in the 4D spacetime the mathematically correct
relations (\ref{E2}) - (\ref{ebv}) are already firmly theoretically founded
and they are known to many physicists. The recent example is, e.g., in [36];
it is only the electric part (the magnetic part is zero there). Similarly,
in the component form these relations are presented, e.g., in [37-41] and in
the basis-free form with AQs in [24, 28, 34, 42]. But, it has to be noted
that from all of them only Oziewicz, see [28] and references to his papers
in it, \emph{exclusively} deals with the abstract, basis-free 4D quantities.
He correctly considers from the outset that in the 4D spacetime such
quantities are physical quantities and not the usual 3D quantities. All
others, starting with Minkowski [16], are not consistent in the use of the
4D electric and magnetic fields. They use the usual 3D fields $\mathbf{E}$
and $\mathbf{B}$ together with the 4D fields considering that the 3D fields
are physically measurable quantities and that their LPET are the
mathematically correct LT, whereas the 4D fields are considered to be only
mathematical, auxiliary, quantities. Minkowski [16] introduced only in Sec.
11.6 the 4D fields and their LT. In other sections he also dealt with the 3D
fields and their LPET. This is explained in detail in [23], which is under
the title: \textquotedblleft Lorentz transformations of the electric and
magnetic fields according to Minkowski.\textquotedblright \bigskip \bigskip

\noindent \textbf{4.\ The proofs that in the ISR the electric field vector
transforms again to the electric field vector\bigskip }

\noindent In the ISR, as proved in Secs. 3.1. and 3.3., the electric field
is properly defined vector on the 4D spacetime and the same holds for the
magnetic field. Hence, \emph{under the LT},\emph{\ }e.g., \emph{the electric
field vector must transform as any other vector transforms}, i.e., \emph{%
again} \emph{to the electric field vector}; \emph{there is no mixing with
the magnetic field vector }$B$. In [20] the same result is obtained for the
electric field as a bivector and for the magnetic field as well. This will
be explicitly shown both for the active LT in 4.1. and for the passive LT in
4.2.\bigskip \medskip

\noindent \textit{4.1. Proof with the coordinate-free quantities, AQs, and
the active LT\bigskip }

\noindent Regarding the correct LT of $E$ and $B$, i.e., MILT, let us start
from the definition with the coordinate-free quantities $E=c^{-1}F\cdot v$
and with \emph{the active LT}. \emph{Mathematically, }as noticed by Oziewicz
[28], \emph{an active LT must act on all tensor fields from which the vector
field }$E$\emph{\ is composed, including an observer's time-like vector field%
}.\ This means that the mathematically correct active LT of $E=c^{-1}F\cdot
v $\ are $E^{\prime }=c^{-1}F^{\prime }\cdot v^{\prime }$; both $F$\ and $v$%
\ are transformed. It was first discovered by Minkowski in Sec. 11.6. in
[16] but with components implicitly taken in the standard basis and, as
already said, reinvented and generalized in terms of 4D GQs in [17-23], see
also Secs. 5. and 6. in [10]. Since this issue is discussed at great length
in [23] and again in [10] we confine our remarks here to a summary of the
conclusions reached in [23]. As explicitly shown, e.g., in [23], in the
geometric algebra formalism \emph{any multivector} $N$ \emph{transforms by
the active LT in the same way,} i.e., as $N\rightarrow N^{\prime }=RN%
\widetilde{R}$, where $R$\ is given by Eq. (10) in [23] (Eq. (39) in [10]);
for boosts in an arbitrary direction the rotor $R$ is $R=(1+\gamma +\gamma
\gamma _{0}\beta )/(2(1+\gamma ))^{1/2}$, where $\gamma =(1-\beta
^{2})^{-1/2}$, the vector $\beta $ is $\beta =\beta s$, $\beta $ on the
r.h.s. of that equation is the scalar velocity in units of $c$ and $s$ is
not the basis vector but any unit space-like vector orthogonal to $\gamma
_{0}$. The reverse $\widetilde{R}$ is defined by the operation of reversion
according to which $\widetilde{AB}=\widetilde{B}\widetilde{A}$, for any
multivectors $A$ and $B$, see Sec. 3. in [23] (Sec. 5. in [10]). Hence, the
vector $E=c^{-1}F\cdot v$\ transforms by \emph{the mathematically correct}
active LT $R$ into $E^{\prime }=RE\widetilde{R}=c^{-1}R(F\cdot v)\widetilde{R%
}=c^{-1}(RF\widetilde{R})\cdot (Rv\widetilde{R})=c^{-1}F^{\prime }v^{\prime
} $. If $v=c\gamma _{0}$ is taken in the expression for $E$\ then $E$
becomes $E=F\cdot \gamma _{0}$ and it transforms according to MILT as in Eq.
(12) in [23], i.e., that \emph{both} $F$ \emph{and} $\gamma _{0}$ are
transformed by the LT, $E=F\cdot \gamma _{0}\longrightarrow E^{\prime
}=R(F\cdot \gamma _{0})\widetilde{R}=(RF\widetilde{R})\cdot (R\gamma _{0}%
\widetilde{R})$. Hence, the explicit form for $E^{\prime }$\ with the
abstract, coordinate-free quantities is given by Eq. (13) in [23], 
\begin{equation}
E^{\prime }=E+\gamma (E\cdot \beta )\{\gamma _{0}-(\gamma /(1+\gamma ))\beta
\}.  \label{aec}
\end{equation}%
In (\ref{aec}) $\beta $ is a vector. That equation is first reported in
[23]. In the ISR Eq. (\ref{aec}) with the 4D vectors replaces Eq. (\ref{JCB}%
) with the 3D vectors; both equations are with geometric quantities but the
3D vectors from (\ref{JCB}) are well defined in the 3D space whereas the 4D
vectors from (\ref{aec}) are well defined in the 4D spacetime. In the
standard basis and for boosts in the direction $x^{1}$ the components of
that $E^{\prime }$\ are 
\begin{equation}
E^{\prime \mu }=(E^{\prime 0}=-\beta \gamma E^{1},\ E^{\prime 1}=\gamma
E^{1},\ E^{\prime 2,3}=E^{2,3}).  \label{LTE}
\end{equation}%
Under the active LT the electric field vector $E=F\cdot \gamma _{0}$ (as a
CBGQ it is $E=E^{\mu }\gamma _{\mu }=0\gamma _{0}+F^{i0}\gamma _{i}$) is
transformed into a new electric field vector $E^{\prime }$, (\ref{aec}).
Note that under the active LT the components are changed, (\ref{LTE}), but
the basis remains unchanged, 
\begin{equation}
E^{\prime \nu }\gamma _{\nu }=-\beta \gamma E^{1}\gamma _{0}+\gamma
E^{1}\gamma _{1}+E^{2}\gamma _{2}+E^{3}\gamma _{3},  \label{eb}
\end{equation}%
Eq. (14) in [23]. \emph{The components} $E^{\mu }$ \emph{transform by the LT
again to the components} $E^{\prime \mu }$ \emph{and there is no mixing with 
}$B^{\mu }$. In general, the LT of the components $E^{\mu }$ (in the $%
\{\gamma _{\mu }\}$ basis) of $E=E^{\mu }\gamma _{\mu }$ are given as 
\begin{equation}
E^{\prime 0}=\gamma (E^{0}-\beta E^{1}),\ E^{\prime 1}=\gamma (E^{1}-\beta
E^{0}),\ E^{\prime 2,3}=E^{2,3},  \label{el}
\end{equation}%
for a boost along the $x^{1}$ axis, i.e., the same LT as for any other 4D
vector.

On the other hand, if in $E=F\cdot \gamma _{0}$ only $F$ is transformed by
the active LT and not $\gamma _{0}$, that is not a mathematically correct
procedure, then the components of that $E_{F}^{\prime }$\ will be denoted as 
$E_{F}^{\prime \mu }$\ and they are 
\begin{equation}
E_{F}^{\prime \mu }=(E_{F}^{\prime 0}=0,E_{F}^{\prime 1}=E^{1},E_{F}^{\prime
2}=\gamma (E^{2}-c\beta B^{3}),E_{F}^{\prime 3}=\gamma (E^{3}+c\beta B^{2})),
\label{ut}
\end{equation}%
Eq. (17) in [23], i.e., (\ref{Em4}) below. \emph{The transformations of the
spatial components (taken in the standard basis) of} $E$ \emph{are exactly
the same as the transformations of }$E_{x,y,z}$ \emph{from }Eq. (11.148)%
\emph{\ in }[2], i.e., as in Eq. (\ref{ee}) below. However, from $E=F\cdot
\gamma _{0}$ it follows that the components of $E$\ are $E^{\mu }=(E^{0}=0$, 
$E^{1}$, $E^{2}$, $E^{3})$. Hence,\emph{\ if only} $F$ \emph{is transformed
by the LT then the temporal components of both} $E$ \emph{and} $%
E_{F}^{\prime }$ \emph{are zero}, $E^{0}=E_{F}^{\prime 0}=0$, which
explicitly reveals that from the ISR viewpoint such transformations are not
the mathematically correct LT; the LT cannot transform $E^{0}=0$ again to $%
E_{F}^{\prime 0}=0$. This proves that from the ISR viewpoint the
transformations (\ref{LTE}) in which \emph{both} $F$ \emph{and} $\gamma _{0}$
are transformed are the mathematically correct LT, the MILT.\bigskip \medskip

\noindent \textit{4.2. Proof with CBGQs and the passive LT\bigskip }

\noindent If $E$\ is written as a CBGQ, i.e., as in (\ref{ebv}),$\ $then we
have to use the passive LT. This is discussed at length in, e.g., [17-22],
[10], but for the sake of completeness and for better understanding we
repeat the short proof from [22]. For example, in the $\gamma _{0}$ - frame $%
E$\ is given as 
\begin{equation}
E=E^{\mu }\gamma _{\mu }=[(1/c)F^{i0}v_{0}]\gamma _{i}=0\gamma
_{0}+E^{i}\gamma _{i}  \label{te1}
\end{equation}%
For boosts in the $\gamma _{1}$\ direction and if \emph{both} $F^{i0}$\ 
\emph{and} $v_{0}$\ are transformed by the LT, i.e., the MILT, then, as for
any other CBGQ, it holds that 
\begin{equation}
E=E^{\mu }\gamma _{\mu }=[(1/c)F^{\prime \mu \nu }v_{\nu }^{\prime }]\gamma
_{\mu }^{\prime }=E^{\prime \mu }\gamma _{\mu }^{\prime },  \label{tec}
\end{equation}%
where, again, the components $E^{\prime \mu }$\ are the same as in (\ref{LTE}%
), see [22]. On the other hand, if only $F^{i0}$ is transformed but not $%
v_{0}$\ then the transformed components $E_{F}^{\prime \mu }$\ are again the
same as in (\ref{ee}) and the same objections as in Sec. 4.1. hold also
here. In addition, it can be easily checked that%
\begin{equation}
E_{F}^{\prime \mu }\gamma _{\mu }^{\prime }\neq E^{\mu }\gamma _{\mu },
\label{fr}
\end{equation}%
which additionaly proves that from the ISR viewpoint the transformations in
which only $F$ is transformed are not the mathematically correct LT and
accordingly the same holds for the transformations given by Eqs. (11.148) ((%
\ref{ut}) here) and (11.149) from [2]. They are the LPET of the components
in the standard basis of the 3D vectors $\mathbf{E}$ and $\mathbf{B}$ that
do not refer to the same 4D quantity. On the other hand, as can be seen from
the above discussion, if $E$\ is written as a CBGQ then, as for any other 4D
CBGQ, e.g., as in (\ref{iks}), it holds that 
\begin{equation}
E=E^{\nu }\gamma _{\nu }=E^{\prime \nu }\gamma _{\nu }^{\prime }=E_{r}^{\nu
}r_{\nu }=E_{r}^{\prime \nu }r_{\nu }^{\prime }.  \label{erc}
\end{equation}%
Again, as in (\ref{iks}), the primed quantities in both bases $\{\gamma
_{\mu }\}$ and $\{r_{\mu }\}$ are the Lorentz transforms of the unprimed
ones.\bigskip \bigskip

\noindent \textbf{5. The comparison of the derivations of the LPET and the
MILT using matrices (the components in the standard basis)}\bigskip \medskip

\noindent \textit{5.1. The electric and magnetic fields as vectors\bigskip }

\noindent For the reader's convenience the same results as in Secs. 3.1. and
3.3. can be obtained explicitly using the matrices. We write the relation $%
E^{\mu }=c^{-1}F^{\mu \nu }v_{\nu }$ in the $\gamma _{0}$ - frame, i.e., for 
$v=c\gamma _{0}$. From the matrix for $F^{\mu \nu }$ and $v_{\nu }=(c,0,0,0)$
one finds $E^{\mu }=(0,F^{10}=E^{1},F^{20}=E^{2},F^{30}=E^{3})$.

Then, \emph{for the LPET only} $F^{\mu \nu }$ \emph{is transformed by the LT
but not the velocity of the observer} $v=c\gamma _{0}$. The Lorentz
transformed $F^{\mu \nu }$ is (symbolically) $F^{\prime }=AF\widetilde{A}$;
here $A$, $F$, ... denote matrices. This relation can be written with
components as $F^{\prime \mu \nu }=A_{\rho }^{\mu }F^{\rho \sigma }%
\widetilde{A}_{\sigma }^{\nu }$. The matrix $A$\ is the boost in the
direction $x^{1}$ (in the standard basis) and it is written in Eq. (\ref{an}%
) below. $A$\ is also given by Eq. (11.98)\ in [2] (with only $\beta
_{1}\neq 0$) and $\widetilde{A}$\ is obtained transposing $A$. The
transformed components $E_{F}^{\prime \mu }$\ are obtained as $E_{F}^{\prime
\mu }=c^{-1}F^{\prime \mu \nu }v_{\nu }$, or explicitly with matrices as 
\begin{equation}
\left[ 
\begin{tabular}{llll}
$0$ & $-F^{\prime 10}$ & $-F^{\prime 20}$ & $-F^{\prime 30}$ \\ 
$E^{1}$ & $0$ & $-F^{\prime 21}$ & $-F^{\prime 31}$ \\ 
$\gamma (E^{2}-\beta cB^{3})$ & $\gamma (-\beta E^{2}+cB^{3})$ & $0$ & $%
-F^{\prime 32}$ \\ 
$\gamma (E^{3}+\beta cB^{2})$ & $\gamma (-\beta E^{3}-cB^{2})$ & $cB^{1}$ & $%
0$%
\end{tabular}%
\right] \cdot \left[ 
\begin{tabular}{l}
$1$ \\ 
$0$ \\ 
$0$ \\ 
$0$%
\end{tabular}%
\right] =\left[ 
\begin{tabular}{l}
$0$ \\ 
$E^{1}$ \\ 
$\gamma (E^{2}-\beta cB^{3})$ \\ 
$\gamma (E^{3}+\beta cB^{2})$%
\end{tabular}%
\right] ,  \label{Em4}
\end{equation}%
where the first matrix is the Lorentz transformed $F^{\mu \nu }$, i.e., $%
F^{\prime \mu \nu }$, and the second matrix is $c^{-1}v^{\mu }=\gamma
_{0}^{\mu }$. The components $E_{F}^{\prime \mu }$\ are already written in
Eq. (\ref{ut}). As seen from (\ref{Em4}) the transformed zeroth component $%
E_{F}^{\prime 0}$ is again $=0$, which shows, as previously stated, that
from the ISR viewpoint, such transformations cannot be the mathematically
correct LT; the LT cannot transform the 4D vector with $E^{0}=0$ into the 4D
vector with $E_{F}^{\prime 0}=0$. Furthermore, it can be simply checked
using (\ref{Em4}) that for the CBGQs holds 
\begin{equation}
E_{F}^{\prime \mu }\gamma _{\mu }^{\prime }\neq E^{\mu }\gamma _{\mu },
\label{em4}
\end{equation}%
where $E_{F}^{\prime \mu }$\ is from (\ref{Em4}).\ This is the same as in (%
\ref{fr}), i.e., it additionally proves that $E_{F}^{\prime \mu }$\ is not
obtained by the mathematically correct LT from $E^{\mu }$. \emph{Under the
mathematically correct LT}, \emph{the MILT}, \emph{\ both} $F^{\mu \nu }$%
\emph{\ and the velocity of the observer }$v=c\gamma _{0}$\emph{\ are
transformed}. Then (symbolically) 
\begin{equation}
E=c^{-1}F\cdot v\longrightarrow E^{\prime }=c^{-1}F^{\prime }\cdot v^{\prime
}=c^{-1}(AF\widetilde{A})(A^{-1}v)=A(c^{-1}Fv)=AE,  \label{ecm}
\end{equation}%
where, here, $E$, $F$, $v$, $A$, $F^{\prime }$, ... denote matrices. Hence, $%
E^{\prime \mu }$\ can be written as 
\begin{equation}
E^{\prime \mu }=c^{-1}F^{\prime \mu \nu }v_{\nu }^{\prime }=c^{-1}(A_{\rho
}^{\mu }F^{\rho \sigma }\widetilde{A}_{\sigma }^{\nu })((A^{-1})_{\nu
}^{\alpha }v_{\alpha })=A_{\rho }^{\mu }(c^{-1}F^{\rho \alpha }v_{\alpha }).
\label{emc}
\end{equation}%
Using the explicit matrices $c^{-1}A^{-1}v$ is given as

\begin{equation}
c^{-1}A^{-1}v=c^{-1}\left[ 
\begin{tabular}{llll}
$\gamma $ & $\beta \gamma $ & $0$ & $0$ \\ 
$\beta \gamma $ & $\gamma $ & $0$ & $0$ \\ 
$0$ & $0$ & $1$ & $0$ \\ 
$0$ & $0$ & $0$ & $1$%
\end{tabular}%
\right] \cdot \left[ 
\begin{tabular}{l}
$c$ \\ 
$0$ \\ 
$0$ \\ 
$0$%
\end{tabular}%
\right] =\left[ 
\begin{tabular}{l}
$\gamma $ \\ 
$\beta \gamma $ \\ 
$0$ \\ 
$0$%
\end{tabular}%
\right]  \label{vc}
\end{equation}%
and $E^{\prime \mu }$\ is $E^{\prime \mu }=c^{-1}F^{\prime \mu \nu }v_{\nu
}^{\prime }$, i.e.,%
\begin{equation}
\left[ 
\begin{tabular}{llll}
$0$ & $-E^{1}$ & $-F^{2^{\prime }0^{\prime }}$ & $-F^{3^{\prime }0^{\prime
}} $ \\ 
$E^{1}$ & $0$ & $-F^{2^{\prime }1^{\prime }}$ & $-F^{3^{\prime }1^{\prime }}$
\\ 
$\gamma (E^{2}-\beta cB^{3})$ & $\gamma (-\beta E^{2}+cB^{3})$ & $0$ & $%
-F^{3^{\prime }2^{\prime }}$ \\ 
$\gamma (E^{3}+\beta cB^{2})$ & $\gamma (-\beta E^{3}-cB^{2})$ & $cB^{1}$ & $%
0$%
\end{tabular}%
\right] \cdot \left[ 
\begin{tabular}{l}
$\gamma $ \\ 
$\beta \gamma $ \\ 
$0$ \\ 
$0$%
\end{tabular}%
\right] =\left[ 
\begin{tabular}{l}
$-\beta \gamma E^{1}$ \\ 
$\gamma E^{1}$ \\ 
$E^{2}$ \\ 
$E^{3}$%
\end{tabular}%
\right] ,  \label{et}
\end{equation}%
where again the first matrix is $F^{\prime \mu \nu }$, as in (\ref{Em4}),\
but the second matrix is the Lorentz transformed 4D velocity of the
observer, i.e., it is given by Eq. (\ref{vc}). Observe that the same result
for $E^{\prime \mu }$\ is obtained from $E^{\prime \mu }=A_{\nu }^{\mu
}E^{\nu }$,%
\begin{equation}
E^{\prime \mu }=A_{\nu }^{\mu }E^{\nu }=\left[ 
\begin{tabular}{llll}
$\gamma $ & $-\beta \gamma $ & $0$ & $0$ \\ 
$-\beta \gamma $ & $\gamma $ & $0$ & $0$ \\ 
$0$ & $0$ & $1$ & $0$ \\ 
$0$ & $0$ & $0$ & $1$%
\end{tabular}%
\right] \cdot \left[ 
\begin{tabular}{l}
$0$ \\ 
$E^{1}$ \\ 
$E^{2}$ \\ 
$E^{3}$%
\end{tabular}%
\right] =\left[ 
\begin{tabular}{l}
$-\beta \gamma E^{1}$ \\ 
$\gamma E^{1}$ \\ 
$E^{2}$ \\ 
$E^{3}$%
\end{tabular}%
\right] .  \label{an}
\end{equation}%
The components $E^{\prime \mu }$\ are the same as in (\ref{LTE}). This
result clearly shows that from the ISR viewpoint the transformations in
which \emph{both} $F$ and the velocity of the observer $v$\ are transformed
are the mathematically correct LT, the MILT; \emph{under such LT, i.e.,
MILT, the electric field 4D vector transforms again only to the electric
field 4D vector as any other 4D vector transforms.}

As an additional proof of that result it can be simply checked using (\ref%
{an}) that for the CBGQs $E^{\nu }\gamma _{\nu }$, $E^{\prime \nu }\gamma
_{\nu }^{\prime }$, ... again holds the relation (\ref{erc}), $E=E^{\nu
}\gamma _{\nu }=E^{\prime \nu }\gamma _{\nu }^{\prime }=E_{r}^{\nu }r_{\nu
}=E_{r}^{\prime \nu }r_{\nu }^{\prime }$, as for any other CBGQ.\bigskip
\medskip

\noindent \textit{5.2. The electric and magnetic fields as bivectors\bigskip 
}

\noindent In [20] the same result about the fundamental difference between
the LPET and the correct LT, MILT, is obtained representing the electric and
magnetic fields by bivectors. The representation by bivectors is used, e.g.,
in [12-15] and they derived the UT in which the components of the
transformed electric field bivector are expressed by the combination of
components of the electric and magnetic field bivectors like in (\ref{ee})
below. In the $\gamma _{0}$ - frame the electric field bivector $\mathbf{E}%
_{H}$ is determined from the electromagnetic field bivector, $\mathbf{E}%
_{H}=(F\cdot \gamma _{0})\gamma _{0}=(1/2)(F-\gamma _{0}F\gamma _{0})$,
which ([20]) can be written as CBGQ in the standard basis $\left\{ \gamma
_{\mu }\right\} $ as $\mathbf{E}_{H}=F^{i0}\gamma _{i}\wedge \gamma _{0}$.
In Sec. 5. in [20] the derivation of the LPET from [12-15] is presented. The
space-time split is made and accordingly the space-space components are zero
for the matrix of the electric field bivector $(\mathbf{E}_{H})^{\mu \nu }$, 
$(\mathbf{E}_{H})^{i0}=F^{i0}=E^{i}$, $(\mathbf{E}_{H})^{ij}=0$.\ Then, in
[12-15], the transformed electric field bivector $\mathbf{E}_{H,at}^{\prime
} $\ is not obtained in the way in which all other multivectors transform ($%
N\rightarrow N^{\prime }=RN\widetilde{R}$, as in Sec. 4.1. here), but it is
obtained that only F is transformed whereas $\gamma _{0}$ is not
transformed, $\mathbf{E}_{H,at}^{\prime }=(1/2)[F^{\prime }-\gamma
_{0}F^{\prime }\gamma _{0}]=(F^{\prime }\cdot \gamma _{0})\gamma _{0}$.\
According to such a treatment from [12-15] the space-space components are
again zero $(\mathbf{E}_{H,at}^{\prime })^{ij}=0$,\ whereas the time-space
components $(\mathbf{E}_{H,at}^{\prime })^{i0}=E_{at.}^{\prime i}$ are given
by the UT (LPET) for the components of the 3D vector $\mathbf{E}$, like in (%
\ref{ee}) below; the transformed components $E_{at.}^{\prime i}$ are
expressed by the mixture of $E^{i}$ and $B^{i}$ components. In [12-15] it is
not noticed that such transformations cannot be the correct LT because the
LT cannot transform the matrix $(\mathbf{E}_{H})^{\mu \nu }$ in which the
space-space components are zero to the matrix $(\mathbf{E}_{H,at}^{\prime
})^{\mu \nu }$\ in which again the space-space components are zero. \emph{%
The space-time split is not a Lorentz covariant procedure.} In Sec. 4. in
[20] the derivation of the correct LT is presented. If the matrix $(\mathbf{E%
}_{H})^{\mu \nu }$ is transformed in the way in which the matrix of any
other bivector transforms under the LT, then the matrix $(\mathbf{E}%
_{H}^{\prime })^{\mu \nu }$\ is obtained in which the space-space components
are different from zero and \emph{the components} $(\mathbf{E}_{H})^{\mu \nu
}$ \emph{transform under the LT again to the components} $(\mathbf{E}%
_{H}^{\prime })^{\mu \nu }$; \emph{there is no mixing with the components of
the matrix of the magnetic field bivector.} In general, as shown in [18, 19]
the electric and magnetic fields can be represented by different algebraic
objects; vectors, bivectors or their combination. \emph{The correct LT
always transform the 4D algebraic object representing the electric field
only to the electric field; there is no mixing with the magnetic field.}%
\bigskip \bigskip

\noindent \textbf{6.\ Briefly avout the field equations and the expressions
for }$T(n)$, $U$ \textbf{and }$S$ \textbf{in terms of vectors }$E$ \textbf{%
and} $B$\textbf{\bigskip }\medskip

\noindent \textit{6.1. A short discussion of the field equations with
vectors\ }$E$ \textit{and} $B$\textit{\bigskip }

\noindent If the decomposition of $F$\ from (\ref{fm}) is introduced into (%
\ref{mf}) then the field equation (\ref{maeb}) is obtained

\begin{align}
\lbrack \partial _{\alpha }(\delta _{\quad \mu \nu }^{\alpha \beta }E^{\mu
}v^{\nu }+\varepsilon ^{\alpha \beta \mu \nu }v_{\mu }cB_{\nu })-& (j^{\beta
}/\varepsilon _{0})]\gamma _{\beta }+  \notag \\
\partial _{\alpha }(\delta _{\quad \mu \nu }^{\alpha \beta }v^{\mu }cB^{\nu
}+\varepsilon ^{\alpha \beta \mu \nu }v_{\mu }E_{\nu })\gamma _{5}\gamma
_{\beta }& =0,  \label{maeb}
\end{align}%
where $E^{\alpha }$ and $B^{\alpha }$ are the basis components in the
standard basis of the 4D vectors $E$ and $B$, $\delta _{\quad \mu \nu
}^{\alpha \beta }=\delta _{\,\,\mu }^{\alpha }\delta _{\,\,\nu }^{\beta
}-\delta _{\,\,\nu }^{\alpha }\delta _{\,\mu }^{\beta }$ and $\gamma _{5}$
is the pseudoscalar in the $\{\gamma _{\mu }\}$ basis. This is Eq. (40) in
[19], but there it is written using some unspecified basis $\left\{ e_{\mu
}\right\} $. The first part in (\ref{maeb}) comes from $\partial \cdot
F=j/\varepsilon _{0}c$ and the second one (the source-free part) comes from $%
\partial \wedge F=0$. As discussed in detail in [19] Eq. (\ref{maeb}) \emph{%
is the relativistically correct, manifestly covariant field equation} that
generalizes the usual Maxwell equations with the 3D fields $\mathbf{E}$ and $%
\mathbf{B}$. It, (\ref{maeb}), can be compared with the usual formulation
with the 3D quantities going to the $\gamma _{0}$-frame in which $v=c\gamma
_{0}$ and Eq. (\ref{sko}) holds. This yields that Eq. (\ref{maeb}) becomes

\begin{align}
(\partial _{k}E^{k}-j^{0}/c\varepsilon _{0})\gamma _{0}+(-\partial
_{0}E^{i}+c\varepsilon ^{ijk0}\partial _{j}B_{k}-j^{i}/c\varepsilon
_{0})\gamma _{i}+&  \notag \\
(-c\partial _{k}B^{k})\gamma _{5}\gamma _{0}+(c\partial
_{0}B^{i}+\varepsilon ^{ijk0}\partial _{j}E_{k})\gamma _{5}\gamma _{i}& =0.
\label{MEC}
\end{align}%
The equation (\ref{MEC}) contains all four usual Maxwell equations in the
component form. The first part (with $\gamma _{\alpha }$) in (\ref{MEC})
contains \emph{two} \emph{Maxwell equations} in the \emph{component form},
the Gauss law for the electric field (the first bracket, with $\gamma _{0}$)
and the Amp\`{e}re-Maxwell law (the second bracket, with $\gamma _{i}$). The
second part (with $\gamma _{5}\gamma _{\alpha }$) contains the \emph{%
component form} of another \emph{two Maxwell equations}, the Gauss law for
the magnetic field (with $\gamma _{5}\gamma _{0}$) and Faraday's law (with $%
\gamma _{5}\gamma _{i}$). As explained in detail in [19] and as seen from (%
\ref{MEC}) in this geometric approach the Amp\`{e}re-Maxwell law and Gauss's
law are inseparably connected in one law and the same happens with Faraday's
law and the law that expresses the absence of magnetic charge. It is not so
in the usual formulation with the 3D\textbf{\ }$\mathbf{E}$\textbf{\ }and $%
\mathbf{B}$.

Observe that the usual component form of the Maxwell equations with the 3D%
\textbf{\ }$\mathbf{E}$\textbf{\ }and $\mathbf{B}$ 
\begin{eqnarray}
\partial _{k}E_{k}-j^{0}/c\varepsilon _{0} &=&0,\quad -\partial
_{0}E_{i}+c\varepsilon _{ikj}\partial _{j}B_{k}-j^{i}/c\varepsilon _{0}=0, 
\notag \\
\partial _{k}B_{k} &=&0,\quad c\partial _{0}B_{i}+\varepsilon _{ikj}\partial
_{j}E_{k}=0  \label{j3}
\end{eqnarray}%
is obtained from the covariant Maxwell equations (\ref{mfc}) using the usual
identifications of six independent components of $F^{\mu \nu }$ with three
components $E_{i}$ and three components $B_{i}$ as in Sec. 7., Eqs. (\ref%
{ieb}) and (\ref{eb2}) below. But, as shown in Sec. 7., such an
identification is meaningless in the $\left\{ r_{\mu }\right\} $ basis,
which means that Maxwell equations (\ref{j3}) do not hold in the $\left\{
r_{\mu }\right\} $ basis. Moreover, the components of the 3D fields from (%
\ref{j3}) transform according to the LPET (\ref{ee}) below and not according
to mathematically correct LT, MILT, (\ref{aec}) - (\ref{el}), which causes,
as explicitly shown in [19], that Eqs. (\ref{j3}) are not covariant under
the LT. On the other hand, contrary to the formulation of the
electromagnetism with the 3D $\mathbf{E}$ and $\mathbf{B}$, \emph{the
formulation with the 4D fields }$E$ \emph{and} $B$, \emph{i.e., with
equation }(\ref{maeb}), \emph{is correct not only in the} $\gamma _{0}$ - 
\emph{frame with the standard basis} $\left\{ \gamma _{\mu }\right\} $ \emph{%
but in all other relatively moving frames and it holds for any permissible
choice of coordinates, i.e., bases}.

This consideration reveals that the 4D fields $E$ and $B$ that transform
like in (\ref{aec}) - (\ref{el}) and the field equation (\ref{maeb}) do not
have the same physical interpretation as the usual 3D fields $\mathbf{E}$
and $\mathbf{B}$ and the usual Maxwell equations (\ref{j3}) except in the $%
\gamma _{0}$ - frame with the $\left\{ \gamma _{\mu }\right\} $ basis in
which $E^{0}=B^{0}=0$.

Here, it is at place a remark about the $\gamma _{0}$ - frame. The
dependence of the relations (\ref{ebv}) and the field equation (\ref{maeb})
on $v$ reflects the arbitrariness in the selection of the $\gamma _{0}$ -
frame, but at the same time this arbitrariness makes that Eqs. (\ref{ebv})
and (\ref{maeb}) are independent of that choice. The $\gamma _{0}$ - frame
can be selected at our disposal depending on the considered problem which
proves that we don't have a kind of \textquotedblleft
preferred\textquotedblright\ frame theory. Some examples will be discussed
in Secs. 10. and 11..

The\ generalization of the field equation for $F$ (\ref{fef}), \textit{i.e.}%
, (\ref{mf}), to a magnetized and polarized moving medium with the
generalized magnetization-polarization bivector $\mathcal{M(}x\mathcal{)}$
is presented in [31]. That generalization is obtained simply replacing $F$
by $F+\mathcal{M}/\varepsilon _{0}$, which yields the primary equations for
the electromagnetism in moving media, Eq. (7) in [31] with AQs and Eq. (8)
in [31] with the CBGQs in the standard basis. It is shown in [31] that if in
equation for $F$\ (\ref{fef}) $j=j^{(C)}+j^{(\mathcal{M})}$ is the total
current density then (\ref{fef}), \textit{i.e.}, (\ref{mf}), \emph{holds
unchanged in moving medium as well}; $j^{(C)}$ is the conduction current
density of the \emph{free} charges and $j^{(\mathcal{M})}=-c\partial \cdot 
\mathcal{M}$ is the magnetization-polarization current density of the \emph{%
bound} charges. $\mathcal{M(}x\mathcal{)}$, in a similar way as for $F$, can
be decomposed into two vectors, the polarization vector $P(x)$ and the
magnetization vector $M(x)$ and the unit time-like vector $u/c$, where the
vector $u$ is identified with bulk 4D velocity vector of the medium in
spacetime; $\mathcal{M}=(1/2)\mathcal{M}^{\mu \nu }\gamma _{\mu }\wedge
\gamma _{\nu }$,$\ \mathcal{M}^{\mu \nu }=(1/c)(P^{\mu }u^{\nu }-P^{\nu
}u^{\mu })+(1/c^{2})\varepsilon ^{\mu \nu \alpha \beta }M_{\alpha }u_{\beta
} $. Hence, the fundamental equations for moving media with the CBGQs in the 
$\{\gamma _{\mu }\}$ basis, Eqs. (29) and (30) in [31], are obtained. Their
sum is the generalization of the field equation (\ref{maeb}) to a magnetized
and polarized moving medium. The equation (29) in [31], i.e., the following
equation, is the part with sources

\begin{equation}
\partial _{\alpha }\{\varepsilon _{0}[\delta _{\quad \mu \nu }^{\alpha \beta
}E^{\mu }v^{\nu }+c\varepsilon ^{\alpha \beta \mu \nu }v_{\mu }B_{\nu
}]+[\delta _{\quad \mu \nu }^{\alpha \beta }P^{\mu }u^{\nu
}+(1/c)\varepsilon ^{\alpha \beta \mu \nu }M_{\mu }u_{\nu }]\}\gamma _{\beta
}=j^{(C)\beta }\gamma _{\beta },  \label{I1}
\end{equation}%
where $\delta _{\quad \mu \nu }^{\alpha \beta }=\delta _{\,\,\mu }^{\alpha
}\delta _{\,\,\nu }^{\beta }-\delta _{\,\,\nu }^{\alpha }\delta _{\,\,\mu
}^{\beta }$. The equation (30) in [31] is the source-free part and it is the
same as the source-free part in (\ref{maeb}). As stated in [31] these
equations are the fundamental equations for moving media and \emph{they
replace all usual Maxwell's equations (with 3D vectors) for moving media}.

Furthermore, in the same way as for vacuum, i.e., as in [33], one can derive
from the field equation the stress-energy vector $T(n)$ for a moving medium
simply replacing $F$ by $F+\mathcal{M}/\varepsilon _{0}$ in Eqs. (26),
(37-47) in [33], i.e., in the equations (\ref{ust}) that are given in Sec.
3. 2. here. The expression for $T(n)$, $T(n)=Un+(1/c)S$, will remain
unchanged, but the energy density $U$ and the Poynting vector $S$ will
change according to the described replacement. This will be important in the
discussion of Abraham-Minkowski controversy.\bigskip \medskip

\noindent \textit{6.2. The expressions for}\textbf{\ }$T(n)$, $U$ \textit{%
and }$S$ \textit{in terms of vectors\ }$E$ \textit{and} $B$\textit{\bigskip }

\noindent Inserting the decomposition of $F$\ (\ref{E2}) into the
coordinate-free expressions for the stress-energy vector $T(n)$, the energy
density $U$, the Poynting vector $S$ (and other quantities) that are found
in [33] and given in Sec. 3.2. here, we can express them in terms of 4D
vectors $E$ and $B$. This most general form for $T(n)$ and the quantities
derived from $T(n)$ will not be presented here, but the special form in
which it is taken that $v=cn$ as in [45]. In that case $T(n)$ takes very
simple form 
\begin{equation}
T(n)=(-\varepsilon _{0}/2)(E^{2}+c^{2}B^{2})n+\varepsilon _{0}c\gamma
_{5}(E\wedge B\wedge n).  \label{tv}
\end{equation}%
\bigskip Again, as before, the first term in (\ref{tv}) ($n-\parallel $ )
yields the energy density $U$ as $U=n\cdot T(n)=(-\varepsilon
_{0}/2)(E^{2}+c^{2}B^{2})$ and the second term ($n-\perp $, i.e., $n\cdot
S=0 $) is $(1/c)$ of the Poynting vector $S.$ The coordinate-free momentum
density $g$ is defined as before $g=(1/c^{2})S$ and the angular-momentum
density is $M=(1/c)T(n)\wedge x=(1/c)U(n\wedge x)+g\wedge x$, where $T(n)$
is given by the relation (\ref{tv}).

All these quantities can be written in some basis $\left\{ e_{\mu }\right\} $%
, e.g., $\{\gamma _{\mu }\}$, $\{r_{\mu }\}$, $\{\gamma _{\mu }^{\prime }\}$%
, etc. bases, as CBGQs. Thus $T(n)$ (\ref{tv}) becomes 
\begin{equation}
T(n)=(-\varepsilon _{0}/2)(E^{\alpha }E_{\alpha }+c^{2}B^{\alpha }B_{\alpha
})n^{\lambda }e_{\lambda }+\varepsilon _{0}c\widetilde{\varepsilon }_{\
\alpha \beta }^{\lambda }E^{\alpha }B^{\beta }e_{\lambda },  \label{tieb}
\end{equation}%
where $\widetilde{\varepsilon }_{\lambda \alpha \beta }=\varepsilon _{\rho
\lambda \alpha \beta }n^{\rho }$ is the totally skew-symmetric Levi-Civita
pseudotensor induced on the hypersurface orthogonal to $n$. The energy
density $U$ in the $\left\{ e_{\mu }\right\} $ basis is determined by the
first term in (\ref{tieb}) $U=(-\varepsilon _{0}/2)(E^{\alpha }E_{\alpha
}+c^{2}B^{\alpha }B_{\alpha }),$ and the Poynting vector $S$ in the $\left\{
e_{\mu }\right\} $ basis is determined by the second term in (\ref{tieb}) as 
$S=c\varepsilon _{0}\widetilde{\varepsilon }_{\ \alpha \beta }^{\lambda
}E^{\alpha }B^{\beta }e_{\lambda }.$ Of course from (\ref{tieb}) one can
easily find $g$ and $M$ in the $\left\{ e_{\mu }\right\} $ basis.

Although we don't need the energy-momentum tensor $T^{\mu \nu }$ (which is
defined in the $\left\{ e_{\mu }\right\} $ basis as $T^{\mu \nu }=T^{\mu
}\cdot e^{\nu }$) we quote here $T^{\mu \nu }$ expressed in terms of
components of 4D vectors $E$ and $B$ in some basis $\left\{ e_{\mu }\right\} 
$ as 
\begin{align}
T^{\mu \nu }& =\varepsilon _{0}[(g^{\mu \nu }/2-n^{\mu }n^{\nu })(E^{\alpha
}E_{\alpha }+c^{2}B^{\alpha }B_{\alpha })-(E^{\mu }E^{\nu }+c^{2}B^{\mu
}B^{\nu })+  \notag \\
& (\varepsilon ^{\mu \alpha \beta \lambda }n_{\lambda }n^{\nu }+\varepsilon
^{\nu \alpha \beta \lambda }n_{\lambda }n^{\mu })cB_{\alpha }E_{\beta }],
\label{t3}
\end{align}%
see also [45] and [46] in which the $\{\gamma _{\mu }\}$ basis is used. It
has to be emphasized once again that, in contrast to all earlier
definitions, our definitions of $T(n)$, $U$, $S$, $g$ and $M$ are the
definitions of the Lorentz invariant quantities.

One can compare these expressions with familiar ones from the 3D space
considering our definitions in the $\gamma _{0}$ - frame (the standard basis 
$\left\{ \gamma _{\mu }\right\} $ and $v=c\gamma _{0}$) and consequently $%
E^{0}=B^{0}=0$. Then $U$ takes the familiar form $U=(-\varepsilon
_{0}/2)(E^{i}E_{i}+c^{2}B^{i}B_{i})$,$\ i=1$, $2$, $3$. (Observe that the
chosen metric is $g_{\mu \nu }=diag(1,-1,-1,-1)$ and that the components $%
E^{i}$ are identified with the components of the 3D $\mathbf{E}$, e.g., $%
E^{1}=E_{x}$,\ which means that $E^{i}E_{i}\ $corresponds to $%
-(E_{x}^{2}+E_{y}^{2}+E_{z}^{2})\ $in terms of the components of the 3D $%
\mathbf{E}$.) Similarly, in the $\gamma _{0}$ - frame, the Poynting vector
becomes the familiar expression $S=\varepsilon _{0}c^{2}\varepsilon _{0\
jk}^{\ i}E^{j}B^{k}\gamma _{i},$ $i$, $j$, $k=1$, $2$, $3$, whence one also
easily finds $g$ and $M$ in the $\gamma _{0}$ - frame. Notice that all
quantities in these expressions are well-defined quantities on the 4D
spacetime. This again nicely illustrates our main idea that \emph{from the
ISR viewpoint in the 4D spacetime the 3D quantities don't exist by
themselves but only as well-defined 4D quantities taken in a particular -
but otherwise arbitrary - inertial frame of reference, here the} $\gamma
_{0} $ \emph{- frame}.\bigskip \bigskip

\noindent \textbf{7. The derivation of the LPET of the 3D} $\mathbf{E}$ 
\textbf{and} $\mathbf{B}$ \textbf{in Jackson [2]\bigskip }

\noindent Einstein's derivation [1] of the LPET of the 3D $\mathbf{E}$ and $%
\mathbf{B}$\ is discussed in [6]. Here, we shall first discuss the
derivation of the LPET from the usual covariant approaches, e.g., from [2].
There the covariant form of the Maxwell equations (\ref{mfc}) is written
with $F^{\alpha \beta }$ and its dual $^{\ast }F^{\alpha \beta }$, where $%
^{\ast }F^{\alpha \beta }=(1/2)\varepsilon ^{\alpha \beta \gamma \delta
}F_{\gamma \delta }$. As already said, in order to get the component form of
the Maxwell equations with the 3D\textbf{\ }$\mathbf{E}$\textbf{\ }and $%
\mathbf{B}$ (\ref{j3}) from equation (\ref{mfc}) one simply makes \emph{the
identification of }the six independent components of $F^{\alpha \beta }$
with six components of the 3D vectors $\mathbf{E}$ and $\mathbf{B}$. These
identifications are

\begin{equation}
E_{i}=F^{i0},\qquad B_{i}=(1/2)\varepsilon _{ijk}F_{kj}  \label{ieb}
\end{equation}%
(the indices $i$, $j$, $k$, $...=1,2,3$), Eq. (11.137) in [2], e.g., $%
E_{x}=E_{1}=F^{10}$. The components of the 3D fields $\mathbf{E}$ and $%
\mathbf{B}$ are written with lowered (generic) subscripts, since they are
not the spatial components of the 4D quantities. This refers to the
third-rank antisymmetric $\varepsilon $ tensor too. The super- and
subscripts are used only on the components of the 4D quantities. The 3D $%
\mathbf{E}$ and $\mathbf{B}$ are \emph{geometric quantities in the 3D space}
and they are constructed from these six independent components of $F^{\mu
\nu }$ and \emph{the unit 3D vectors }$\mathbf{i}$, $\mathbf{j}$, $\mathbf{k}
$, e.g., $\mathbf{E=}F^{10}\mathbf{i}+F^{20}\mathbf{j}+F^{30}\mathbf{k}$.
Observe that $F^{\alpha \beta }$ is not a tensor\ since $F^{\alpha \beta }$
are only components implicitly taken in the standard basis. The components
are coordinate quantities and they do not contain the whole information
about the physical quantity, since a basis of the spacetime is not included.
In the covariant approaches, e.g., [2], one transforms by the passive LT the
covariant Maxwell equations (\ref{mfc}) and finds 
\begin{equation}
\partial _{\alpha }^{\prime }F^{\prime a\beta }=j^{\prime \beta
}/\varepsilon _{0}c,\quad \partial _{\alpha }^{\prime }\ ^{\ast }F^{\prime
\alpha \beta }=0.  \label{cme}
\end{equation}%
Then, it is supposed that the same identification of the components as in
equation (\ref{ieb}) holds for a relatively moving inertial frame $S^{\prime
}$, i.e., for the transformed components $E_{i}^{\prime }$ and $%
B_{i}^{\prime }$

\begin{equation}
E_{i}^{\prime }=F^{\prime i0},\quad B_{i}^{\prime }=(1/2c)\varepsilon
_{ijk}F_{kj}^{\prime }.  \label{eb2}
\end{equation}%
The same remark about the (generic) subscripts holds also here. The
components $F^{\alpha \beta }$ transform under the LT as, e.g.,%
\begin{equation}
F^{\prime 10}=F^{10},\ F^{\prime 20}=\gamma (F^{20}-\beta F^{21}),\
F^{\prime 30}=\gamma (F^{30}-\beta F^{31}),  \label{fe}
\end{equation}%
which yields (by equations (\ref{ieb}) and (\ref{eb2})) that 
\begin{equation}
E_{1}^{\prime }=E_{1},\ E_{2}^{\prime }=\gamma (E_{2}-\beta cB_{3}),\
E_{3}^{\prime }=\gamma (E_{3}+\beta cB_{2}),  \label{ee}
\end{equation}%
what is equation (11.148) in [2]. Thus, in such approaches, e.g., [2], \emph{%
the LPET of the components of} $\mathbf{E}$ \emph{and} $\mathbf{B}$ \emph{%
are derived assuming that they transform under the LT as the components of} $%
F^{\alpha \beta }$ \emph{transform}.

However, from the mathematical viewpoint, i.e., from the ISR viewpoint,
there are several objections to the mathematical correctness of such a
procedure. Some of them are the following:

1) As seen, e.g., from Sec. 3.1 in [10], such an identification of the
components of $\mathbf{E}$ and $\mathbf{B}$ with the components of $%
F^{\alpha \beta }$ is synchronization dependent and, particularly, it is
meaningless in the \textquotedblleft radio,\textquotedblright\
\textquotedblleft r\textquotedblright\ synchronization, i.e., in the $%
\left\{ r_{\mu }\right\} $ basis, see [6] and below.

2) The 3D vectors $\mathbf{E}$, $\mathbf{B}$ and $\mathbf{E}^{\prime }$, $%
\mathbf{B}^{\prime }$ are constructed in both frames in the same way, i.e.,
multiplying the components, e.g., $E_{x,y,z}$ and$\ E_{x,y,z}^{\prime }$ by
the unit 3D vectors $\mathbf{i}$, $\mathbf{j}$, $\mathbf{k}$ and $\mathbf{i}%
^{\prime }$, $\mathbf{j}^{\prime }$, $\mathbf{k}^{\prime }$, respectively.
This procedure gives the LPET of the 3D vectors $\mathbf{E}$ and $\mathbf{B}$%
, Eq. (11.149) in [2]. But, as seen from (\ref{fc}) below, \emph{the
components} $F^{\alpha \beta }$ \emph{are multiplied by the bivector basis} $%
\gamma _{\alpha }\wedge \gamma _{\beta }$ \emph{and not by the unit 3D
vectors. In the 4D spacetime the unit 3D vectors are ill-defined algebraic
quantities and there are no LT, or some other transformations, that
transform the unit 3D vectors} $\mathbf{i}$, $\mathbf{j}$, $\mathbf{k}$ 
\emph{into the unit 3D vectors} $\mathbf{i}^{\prime }$, $\mathbf{j}^{\prime
} $, $\mathbf{k}^{\prime }$.

3) As mentioned above (the objection 1) the identification of the components
of $\mathbf{E}$ and $\mathbf{B}$ with the components of $F^{\alpha \beta }$,
(\ref{ieb}), is synchronization dependent. If the components $F^{\alpha
\beta }$ of $F$ are transformed by the transformation matrix $R_{\;\nu
}^{\mu }$ to the $\{r_{\mu }\}$ basis, then it is obtained that, e.g.,%
\begin{equation}
F_{r}^{10}=F^{10}-F^{12}-F^{13}.  \label{are}
\end{equation}%
Hence, as shown, e.g., in [6] and [10], in the $\left\{ r_{\mu }\right\} $
basis the identification $E_{1r}=F_{r}^{10}$, as in (\ref{ieb}), yields that
the component $E_{1r}$ is expressed as the combination of $E_{i}$ and $B_{i}$
components from the $\left\{ \gamma _{\mu }\right\} $ basis

\begin{equation}
E_{1r}=F_{r}^{10},\quad E_{1r}=E_{1}+cB_{3}-cB_{2}.  \label{FEr}
\end{equation}%
This means that\textit{\ if the \textquotedblleft r\textquotedblright\
synchronization is used then it is not possible to make the usual
identifications} (\ref{ieb}) \textit{and} (\ref{eb2}).

4) As discussed in [33] and in Sec. 3.2. here in the 4D geometric approach,
i.e., in the ISR, the primary quantity for the whole electromagnetism is a
physically measurable quantity, the bivector field $F=(1/2)F^{\mu \nu
}\gamma _{\mu }\wedge \gamma _{\nu }$, where $\gamma _{\mu }\wedge \gamma
_{\nu }$ is the bivector basis and the basis components $F^{\mu \nu }$\ are
determined as $F^{\mu \nu }=\gamma ^{\nu }\cdot (\gamma ^{\mu }\cdot
F)=(\gamma ^{\nu }\wedge \gamma ^{\mu })\cdot F$. In the same way as for any
other CBGQ it holds that bivector $F$ \emph{is the same 4D quantity} for
relatively moving inertial observers and for all bases chosen by them, e.g.,

\begin{equation}
F=(1/2)F^{\mu \nu }\gamma _{\mu }\wedge \gamma _{\nu }=(1/2)F_{r}^{\mu \nu
}r_{\mu }\wedge r_{\nu }=(1/2)F^{\prime \mu \nu }\gamma _{\mu }^{\prime
}\wedge \gamma _{\nu }^{\prime }=(1/2)F_{r}^{\prime \mu \nu }r_{\mu
}^{\prime }\wedge r_{\nu }^{\prime },  \label{fc}
\end{equation}%
where, as in (\ref{erc}), the primed quantities in both bases $\{\gamma
_{\mu }\}$ and $\{r_{\mu }\}$ are the Lorentz transforms of the unprimed
ones. \emph{Only the whole} $F$\ \emph{from} (\ref{fc}) \emph{is a
mathematically correctly defined quantity} and it does have a definite
physical reality. The components $F^{i0}$, or $F^{ij}$ (implicitly
determined in the standard basis $\{\gamma _{\mu }\}$), if taken alone, are
not properly defined physical quantities in the 4D spacetime. The
transformations of these components, e.g., Eq. (\ref{fe}), which are
extracted from the LT\ of the whole properly defined physical quantity $%
F=(1/2)F^{\alpha \beta }\gamma _{\alpha }\wedge \gamma _{\beta }$, are not
the mathematically correct LT. They do not refer to \emph{the same }4\emph{D}
\emph{quantity} for relatively moving observers. Hence, from the ISR
viewpoint, the determination of $\mathbf{E}$ and $\mathbf{B}$ by the
components $F^{i0}$ and $F^{ij}$, respectively, as the quantities that do
not depend on the 4D velocity of the observer is not mathematically correct.
In contrast to it, the determination of vectors $E$ and $B$ relative to the
observer by the decomposition of $F$, i.e., by Eqs. (\ref{E2}) and (\ref{E1}%
) with coordinate-free quantities, or (\ref{fm}) and (\ref{ebv}) with the
CBGQs is mathematically correct. Every antisymmetric tensor of the second
rank (as a geometric quantity) can be decomposed into two vectors and a unit
timelike vector, in this case, $v/c$. The components are coordinate
quantities and they are only a part of the representation in some basis of
an abstract, coordinate-free bivector $F$.

5) In addition, it is worth mentioning that in the usual covariant
approaches, e.g., [2], the components $F^{\alpha \beta }$ are defined in
terms of a 4D vector potential $A^{\alpha }=(\Phi ,\mathbf{A})$, Eq.
(11.132) in [2], as $F^{\alpha \beta }=\partial ^{\alpha }A^{\beta
}-\partial ^{\beta }A^{\alpha }$, Eq. (11.136) in [2]. The 3D fields $%
\mathbf{E}$ and $\mathbf{B}$ are determined in terms of the potentials by
Eq. (11.134) in [2], which, together with Eq. (11.136) in [2], leads to Eq.
(11.137) in [2] in which, as already stated, the components $F^{\alpha \beta
}$ are expressed in terms of the components of the 3D vectors $\mathbf{E}$
and $\mathbf{B}$. According to that procedure from [2]\ the 4D vector
potential $A^{\alpha }$ (gauge dependent and thus unmeasurable quantity) is
considered to be the primary quantity which determines the measurable
quantities, the electric and magnetic fields and also $F^{\alpha \beta }$.
Observe that, contrary to the assertions from [2], $A^{\alpha }$ is not a 4D
vector. $A^{\alpha }$ are only components implicitly taken in the standard
basis of the 4D vector $A=A^{\mu }\gamma _{\mu }$. In the 4D spacetime only
the whole 4D potential $A=A^{\mu }\gamma _{\mu }=A_{r}^{\mu }r_{\mu }$ is a
well-defined quantity,\ whereas it is not the case with the usual scalar
potential $\Phi $\ and the 3D vector potential $\mathbf{A}$ in which the
components $A_{x,y,z}$ are multiplied by the unit 3D vectors $\mathbf{i}$, $%
\mathbf{j}$, $\mathbf{k}$ and not by the properly defined unit 4D vectors $%
\gamma _{\mu }$.

Similar derivations of the LPET of the 3D $\mathbf{E}$ and $\mathbf{B}$\ are
given in many other textbooks, e.g., in \textquotedblleft R. P. Feynman, R.
B. Leighton and M. Sands, \textit{The Feynman Lectures on Physics} Volume II
(Addison-Wesley, Reading, 1964)\textquotedblright\ in Sec. 26-3 under the
title \textquotedblleft Relativistic transformation of the
fields,\textquotedblright\ in \textquotedblleft L. D. Landau and E. M.
Lifshitz, \textit{The Classical Theory of Fields}, 4th ed. (Pergamon, New
York, 1975 )\textquotedblright\ in Sec. 24 under the title \textquotedblleft
Lorentz transformation of the field\textquotedblright , etc.. All objections
1) - 5) from this section hold in the same measure for the mentioned
well-known textbooks.

In Sec. 12.3.2 in [25] under the title \textquotedblleft How the Fields
Transform\textquotedblright\ the LPET of the 3D $\mathbf{E}$ and $\mathbf{B}$
(only components implicitly taken in the standard basis) Eq. (12. 109) in
[25], are derived using the Lorentz contraction, the time dilation and the
3D fields. But, as discussed in Sec. 1. and in Appendix here, the Lorentz
contraction and the time dilation are ill-defined in the 4D spacetime; they
are synchronization dependent and consequently they are not intrinsic
relativistic effects. That derivation from [25] is the same as in
\textquotedblleft E.M. Purcell, \textit{Electricity and Magnetism},\textit{\ 
}2nd ed. (McGraw-Hill, New York, 1985)\textquotedblright . The derivation of
the LPET of the 3D $\mathbf{E}$ and $\mathbf{B}$ from Purcell's textbook is
discussed at great length and objected from the ISR viewpoint in Sec. 4.3.
in [6] and will not be repeated here.\bigskip \bigskip

\noindent \textbf{8. The derivation of the LPET of the 3D} $\mathbf{E}$ 
\textbf{and} $\mathbf{B}$ \textbf{in Blanford and\ Thorne }[\textbf{24}]%
\textbf{\bigskip }

\noindent As mentioned in the Introduction \emph{the nature of electric and
magnetic fields is discussed}\ by Blandford and Thorne (BT) in Sec. 1.10 in
[24]. There, it is concluded that \emph{these fields are the 4D fields.} If
one applies the LT to BT's equation (1.109) (it is our Eq. (\ref{ebv})),
e.g., to the electric field 4D vector then, as discussed above, \emph{both} $%
F^{\alpha \beta }$ \emph{and} $w_{\beta }$ (their $w$ is our $v$) have to be
transformed. The equation (\ref{LTE}) would be obtained and Eq. (\ref{erc})
would hold. This is \emph{not} noticed by BT, [24], and they believe as all
others that their Eq. (1.113) with the 3D vectors (the same as Eq. (11.149)
in [2], i.e., Eq. (\ref{JCB}) here) is the mathematically correct
\textquotedblleft \textit{Relationship Between Fields Measured by Different
Observers}.\textquotedblright\ Thus, although they deal with 4D GQs they
still consider that in the 4D spacetime, in the same way as in the 3D space,
the 3D vectors are the physical quantities, whereas the 4D quantities are
considered to be only mathematical, auxiliary, quantities. This is visible
in the treatment of the Lorentz force in [24]. In the usual formulations the
physical meaning of 3D vectors $\mathbf{E}$ and $\mathbf{B}$ is determined
by the Lorentz force as a 3D vector $\mathbf{F}_{L}\mathbf{=}q\mathbf{E}+q%
\mathbf{u}\times \mathbf{B}$\ and by Newton's second law $\mathbf{F}=d%
\mathbf{p}/dt$, $\mathbf{p=}m\gamma _{u}\mathbf{u}$. BT start with the
correct equation (1.106) ($dp^{\mu }/d\tau =(q/c)F^{\mu \nu }u_{\nu }$, our
notation), but then instead of to use the decomposition of $F^{\mu \nu }$,
their Eq. (1.110), our Eq. (\ref{fm}), but only components $F^{\mu \nu }$, 
\emph{they deal with the usual identification of the components (in the
standard basis) of} $F^{\mu \nu }$ \emph{with} \emph{the components of the
3D vectors} $\mathbf{E}$ \emph{and} $\mathbf{B}$, their Eq. (1.107), our Eq.
(\ref{ieb}), which, as discussed above, is synchronization dependent and
even meaningless in the $\{r_{\mu }\}$ basis, see Eqs. (\ref{are}) and (\ref%
{FEr}). Finally they get \textquotedblleft the familiar Lorentz-force
form\textquotedblright\ in terms of \emph{the 3D vectors} $\mathbf{E}$ and $%
\mathbf{B}$, their Eq. (1.108). Thus, the same as in the usual approaches.
It is interesting that Thorne and Blandford (TB) applied the same
consideration about the Lorentz force as above in their recent very good
textbook [47] that is written in geometric terms.

However, in the 4D spacetime, as mentioned above, the Lorentz force $K_{L}$\
is given by Eq. (\ref{LF}) in terms of $F$ and $u$. Using the decomposition
of $F$ (\ref{E2}) the Lorentz force $K_{L}$\ becomes 
\begin{equation}
K_{L}=(q/c)\left[ (1/c)E\wedge v+(IB)\cdot v\right] \cdot u,  \label{lfa}
\end{equation}%
where $u$ is the velocity vector of a charge $q$ (it is defined to be the
tangent to its world line). Note that \emph{there are two velocity vectors in%
} $K_{L}$\ \emph{if it is expressed in terms of fields} $E$ \emph{and} $B$,
because $E$ and $B$ are determined relative to the observer with velocity
vector $v$. If $K_{L}$\ is represented as a CBGQ in the standard basis it is 
\begin{equation}
K_{L}=K_{L}^{\mu }\gamma _{\mu }=(q/c)F^{\mu \nu }u_{\nu }\gamma _{\mu
}=(q/c)\{[(1/c)(E^{\mu }v^{\nu }-E^{\nu }v^{\mu })+\varepsilon ^{\lambda \mu
\nu \rho }v_{\lambda }B_{\rho }]u_{\nu }\}\gamma _{\mu },  \label{lf}
\end{equation}%
where $F^{\mu \nu }$\ is from Eq. (\ref{fm}). In contrast to the usual
expression for the Lorentz force with the 3D fields $\mathbf{E}$ and $%
\mathbf{B}$, $\mathbf{F}_{L}\mathbf{=}q\mathbf{E}+q\mathbf{u}\times \mathbf{B%
}$, the Lorentz force with the 4D fields $E$ and $B$ (\ref{lfa}) or (\ref{lf}%
) contains not only the 4D velocity $u$ of a charge $q$\ but also the 4D
velocity $v$ of the observer who measures 4D fields. It can be simply
checked that for $K_{L}^{\mu }\gamma _{\mu }$\ the following relation holds 
\begin{equation}
K_{L}=K_{L}^{\mu }\gamma _{\mu }=K_{L}^{\prime \mu }\gamma _{\mu }^{\prime
}=K_{Lr}^{\mu }r_{\nu }=K_{Lr}^{\prime \mu }r_{\nu }^{\prime }  \label{klc}
\end{equation}%
as for any other 4D CBGQ. \emph{In the 4D spacetime, the physical meaning of}
$E^{\mu }$ \emph{and} $B^{\mu }$ \emph{is determined by the Lorentz force} $%
K_{L}$ (\ref{lfa}), i.e., $\ K_{L}^{\mu }\gamma _{\mu }$\ (\ref{lf}) \emph{%
and by the 4D expression for Newton's second law} 
\begin{equation}
K_{L}^{\mu }\gamma _{\mu }=(dp^{\mu }/d\tau )\gamma _{\mu },\quad p^{\mu
}=mu^{\mu },  \label{pm}
\end{equation}%
$p^{\mu }$ is the proper momentum (components) and $\tau $\ is the proper
time. \emph{All components} $E^{\mu }$ \emph{and} $B^{\mu }$, \emph{thus} $%
E^{0}$ \emph{and} $B^{0}$ \emph{as well, are equally well physical and
measurable quantities by means of the mentioned} $K_{L}^{\mu }$\ (\ref{lf}) 
\emph{and the 4D expression for Newton's second law} (\ref{pm}). Hence, in
the 4D spacetime, contrary to the assertion from [24, 47], the use of the
mathematically correct 4D GQs as in (\ref{lfa}) or (\ref{lf}) cannot lead to
\textquotedblleft the familiar Lorentz-force form.\textquotedblright

Furthermore, BT in [24] (and TB in [47]), state: \textquotedblleft Only
after making such an observer-dependent \textquotedblleft 3+1
split\textquotedblright\ of spacetime into space plus time do the electric
field and magnetic field come into existence as separate
entities.\textquotedblright\ But, as shown above, in the 4D spacetime
\textquotedblleft 3+1 split\textquotedblright\ is ill-defined. It does not
hold in the $\{r_{\mu }\}$ basis and even in the $\{\gamma _{\mu }\}$ basis 
\emph{it is not a Lorentz covariant procedure}, i.e., the 3-surface of
simultaneity for one observer (with 4D velocity $w$) cannot be transformed
by the LT into the 3-surface of simultaneity for a relatively moving
inertial observer (with 4D velocity $w^{\prime }$). If for one observer$\
w^{\mu }=(1,0,0,0)\ $then for a relatively moving inertial observer it holds
that $w^{\prime \mu }=(\gamma ,-\beta \gamma ,0,0)$). Hence, it cannot be
mathematically correct that \emph{both} $E_{w}^{0}=0$ and\ $E_{w^{\prime
}}^{0}=0$, but it is necessary $E_{w^{\prime }}^{0}\neq 0$, as in (\ref{LTE}%
) or (\ref{an}). This means that their, [24], Eq. (1.107), our Eq. (\ref{ieb}%
), is not correct. It does not follow from Eq. (1.109), our Eq. (\ref{ebv})
(without unit 4D vectors). Also, Eq. (1.113) cannot be obtained by a
mathematically correct procedure from Eq. (1.110). Simply, in the 4D
spacetime there is no room for the 3D quantities; an independent physical
reality has to be consistently attributed to the 4D GQs and not to the usual
3D quantities. Obviously, an important statement from Chapter 1 in [24] that
is already mentioned above: \textquotedblleft \emph{We shall state physical
laws, e.g. the Lorentz force law, as geometric, coordinate-free
relationships between these geometric, coordinate free quantities,}%
\textquotedblright\ has to be changed in this way: \emph{In the 4D spacetime
physical laws, e.g. the Lorentz force law, are geometric, coordinate-free
relationships between the 4D geometric, coordinate free quantities}.

The 3D fields $\mathbf{E}$ and $\mathbf{B}$ and the Lorentz force $\mathbf{F}%
_{L}$ ($\mathbf{F}_{L}=q\mathbf{E}+q\mathbf{u}\times \mathbf{B}$) are also
geometric quantities but in the 3D space, which means that they do not have
well-defined mathematical and physical meaning in the 4D spacetime.

In addition, BT in [24] (TB in [47]) consider, as almost the whole physics
community, that the Lorentz contraction and the time dilation are the
intrinsic relativistic effects. However, as already mentioned, in [6], [7]
and in Appendix here, it is exactly proved that such an opinion is not
correct since both the Lorentz contraction and the time dilation are
ill-defined in the 4D spacetime. Instead of them the 4D GQs, the position 4D
vector, the distance 4D vector between two events and the spacetime length
have to be used, since they are properly defined quantities in the 4D
spacetime\bigskip \bigskip

\noindent \textbf{9. Additional remarks\ about the 4D Lorentz force}\bigskip

\noindent Here, it is at place to give some additional comments about the
Lorentz force $K_{L}$ (\ref{lfa}) or (\ref{lf}) as a 4D GQ. It is visible
from (\ref{lfa}) or (\ref{lf}) that the Lorentz force ascribed by an
observer comoving with a charge, $u=v$, i.e., if the charge and the observer
world lines coincide, then $K_{L}$ is purely electric, $K_{L}=qE$. In the
general case when $u$ is different from $v$, i.e. when the charge and the
observer have distinct world lines, $K_{L}$ (\ref{lfa}) or (\ref{lf}) can be
written in terms of $E$ and $B$ as a sum of the $v$ - orthogonal part, $%
K_{L\perp }$ ($K_{L\perp }\cdot v=0$) and $v$ - parallel part, $%
K_{L\parallel }$ ($K_{L\parallel }\wedge v=0$). As the CBGQs they are%
\begin{eqnarray}
K_{L} &=&K_{L\perp }+K_{L\parallel },\quad K_{L\perp }=(q/c^{2})[(v^{\nu
}u_{\nu })E^{\mu }+\varepsilon ^{\lambda \mu \nu \rho }v_{\lambda }u_{\nu
}cB_{\rho }]\gamma _{\mu },  \notag \\
K_{L\parallel } &=&(q/c^{2})[-(E^{\nu }u_{\nu })v^{\mu }]\gamma _{\mu }.
\label{kop}
\end{eqnarray}%
Speaking in terms of the prerelativistic notions one can say that in the
approach with the vectors $E$ and $B$ the $v$ - orthogonal part, $K_{L\perp
} $, from (\ref{kop}) plays the role of the usual Lorentz force lying on the
3D hypersurface orthogonal to $v$, whereas $K_{L\parallel }$ from (\ref{kop}%
) is related to the work done by the field on the charge. This can be seen
specifying (\ref{kop}) to the $\gamma _{0}$ - frame, $v=c\gamma _{0}$, in
which $E^{0}=B^{0}=0$. In the $\gamma _{0}$ - frame it is possible to
compare the 4D vector $K_{L}$\ with the usual 3D Lorentz force, $\mathbf{F}%
_{L}\mathbf{=}q\mathbf{E}+q\mathbf{u}\times \mathbf{B}$, which yields%
\begin{eqnarray}
K_{L}^{0}\gamma _{0} &=&K_{L\parallel }^{0}\gamma
_{0}=-(q/c)E^{i}u_{i}\gamma _{0},\quad K_{L\perp }^{0}=0,  \notag \\
K_{L}^{i}\gamma _{i} &=&K_{L\perp }^{i}\gamma _{i}=q((E^{i}+\varepsilon
^{0ijk}u_{j}B_{k})\gamma _{i},\quad K_{L\parallel }^{i}\gamma _{i}=0
\label{koi}
\end{eqnarray}%
It is visible from (\ref{koi}) that $K_{L}^{0}$\ is completely determined by 
$K_{L\parallel }$,\ whereas the spatial components $K_{L}^{i}$\ are
determined by $K_{L\perp }$.\ However, as already mentioned several times,
in this 4D geometric approach, i.e., in the ISR, \emph{only both parts taken
together, i.e., the whole }$K_{L}=K_{L\perp }+K_{L\parallel }$\emph{\ does
have a definite physical meaning and it defines the 4D Lorentz force both in
the theory and in experiments. }

In Sec. 2.5 in [33], under the title \textquotedblleft The Lorentz force and
the motion of charged particle in the electromagnetic field $F$%
\textquotedblright\ the definition of $K_{L}$ in terms of $F$ is exclusively
used ($K_{L}=(q/c)F\cdot u$) without introducing the electric and magnetic
fields. Observe that the 4D GQs $K$ ($K_{L}$), $p$, $u$ transform in the
same way, like any other 4D vector, i.e., according to the LT and not
according to the awkward usual transformations of the 3D force $\mathbf{F}$,
e.g., Eqs. (12.65) - (12.67) in [25], and the 3D momentum $\mathbf{p}$,
i.e., the 3D velocity $\mathbf{u}$. In [48], under the title
\textquotedblleft Four Dimensional Geometric Quantities versus the Usual
Three-Dimensional Quantities: The Resolution of Jackson's
Paradox,\textquotedblright\ it is shown that only with the use of the 4D
Lorentz force (\ref{lfa}), (\ref{lf}) or (\ref{kop}), the torque bivector $%
N=(1/2)N^{\mu \nu }\gamma _{\mu }\wedge \gamma _{\nu }$, $N^{\mu \nu
}=x^{\mu }K_{L}^{\nu }-x^{\nu }K_{L}^{\mu }$ and the angular momentum
bivector $M=(1/2)M^{\mu \nu }\gamma _{\mu }\wedge \gamma _{\nu }$, $M^{\mu
\nu }=m(x^{\mu }u^{\nu }-x^{\nu }u^{\mu })$ there is no apparent
electrodynamic paradox with the torque and that the principle of relativity
is naturally satisfied. The paper [49] is a simpler version of the paper
[48]. The mentioned paradox is described in [50]\ and it consists in the
fact that there is a 3D torque $\mathbf{N}$ and thus $d\mathbf{L}/dt$ ($%
\mathbf{N}=d\mathbf{L}/dt$) in one inertial frame, but no 3D angular
momentum $\mathbf{L}^{\prime }$ and no 3D torque $\mathbf{N}^{\prime }$ in
another relatively moving inertial frame. Similar electrodynamic paradoxes
with the 3D torque appear in the Trouton-Noble paradox, see, e.g., [51] and
the \textquotedblleft charge-magnet paradox\textquotedblright\ [52 - 54].
Using the above mentioned 4D GQs, 4D Lorentz force, the torque and angular
momentum bivectors it is explicitly shown in [55], [33] for the
Trouton-Noble paradox and [56], [10] for Mansuripur's paradox that there is
no paradox and consequently there is no need for some \textquotedblleft
resolutions\textquotedblright\ of the paradoxes, e.g., by the introduction
of the Einstein-Laub force, [52 - 54], or by the introduction of some
\textquotedblleft hidden\textquotedblright\ quantities, e.g., [57 -
61].\bigskip \bigskip

\noindent \textbf{10. The electromagnetic field of a point charge in uniform
motion}\bigskip

\noindent It is worth mentioning that the majority of physicists consider
that if the electric field would be transformed by the LT again into the
electric field as in (\ref{LTE}) then it would imply that moving electrons
produce no magnetic field.\ In Sec. 5.6 in [56] the electromagnetic field of
a point charge in uniform motion is treated in detail. There it is shown
that the formulation of that problem with the 4D fields and their MILT (\ref%
{aec}), (\ref{LTE}) is mathematically completely correct but its physical
interpretation is different than in the usual formulation with the 3D fields
and their LPET. The consideration presented in 5.6.2 - 5.6.2.2 in [56]
explicitly shows that the formulation with the 4D fields that transform
according to the MILT (\ref{aec}), (\ref{LTE}) simply explains the existence
of the electric \emph{and magnetic fields} for a moving electron.\bigskip
\medskip

\noindent \textit{10.1. The bivector field}\textbf{\ }$F$\textit{\bigskip }

\noindent Here we shall briefly quote the main results from [56]. \emph{In
the 4D formulation the primary quantity is the bivector field} $F$. The
expression for $F$ for an arbitrary motion of a point charge is given in
[33] by Eqs. (10) (coordinate-free quantities) and (11) (CBGQs).
Particularly, for a charge $Q$ moving with constant 4D velocity vector $u$, $%
F$ is given by Eq. (12) in [33] (coordinate-free quantities), i.e., Eq. (65)
in [56]

\begin{equation}
F(x)=G(x\wedge (u/c)),\quad G=kQ/\left\vert x\wedge (u/c)\right\vert ^{3},
\label{fcf}
\end{equation}%
where $k=1/4\pi \varepsilon _{0}$. $G$ is a number, a Lorentz scalar. The
geometric character of $F$ is contained in $x\wedge (u/c)$. If that $F$\ is
written as a CBGQ in the standard basis it is

\begin{equation}
F=(1/2)F^{\mu \nu }\gamma _{\mu }\wedge \gamma ;F^{\mu \nu }=G(1/c)(x^{\mu
}u^{\nu }-x^{\nu }u^{\mu }),G=kQ/[(x^{\mu }u_{\mu })^{2}-c^{2}x^{\mu }x_{\mu
}]^{3/2}.  \label{fmn}
\end{equation}%
In order to find the explicit expression for $F$ from (\ref{fmn}) in the $%
S^{\prime }$ frame in which the charge $Q$ is at rest one has simply to put
into (\ref{fmn})\ that $u=c\gamma _{0}^{\prime }$ with $\gamma _{0}^{\prime
\mu }=(1,0,0,0)$. Then, $F=(1/2)F^{\prime \mu \nu }\gamma _{\mu }^{\prime
}\wedge \gamma _{\nu }^{\prime }$ and

\begin{equation}
F=F^{\prime i0}(\gamma _{i}^{\prime }\wedge \gamma _{0}^{\prime
})=Gx^{\prime i}(\gamma _{i}^{\prime }\wedge \gamma _{0}^{\prime }),\quad
G=kQ/(x^{\prime i}x_{i}^{\prime })^{3/2}.  \label{q1}
\end{equation}%
In $S^{\prime }$ and in the standard basis, the basis components $F^{\prime
\mu \nu }$ of the bivector $F$ are obtained from (\ref{fmn}) and they are: 
\begin{equation}
F^{\prime i0}=-F^{\prime 0i}=kQx^{\prime i}/(x^{\prime i}x_{i}^{\prime
})^{3/2},\quad F^{\prime ij}=0.  \label{fci}
\end{equation}%
In the charge's rest frame there are only components $F^{\prime i0}$, which
are the same as the usual components of the 3D electric field $\mathbf{E}$
for a charge at rest.

In the same way we find the expression for $F$ (\ref{fmn}) in the $S$ frame
in which the charge $Q$ is moving, i.e., $u=u^{\mu }\gamma _{\mu }$ with $%
u^{\mu }/c=(\gamma ,\gamma \beta ,0,0)$. Then 
\begin{gather}
F=G\gamma \lbrack (x^{1}-\beta x^{0})(\gamma _{1}\wedge \gamma
_{0})+x^{2}(\gamma _{2}\wedge \gamma _{0})+x^{3}(\gamma _{3}\wedge \gamma
_{0})  \notag \\
-\beta x^{2}(\gamma _{1}\wedge \gamma _{2})-\beta x^{3}(\gamma _{1}\wedge
\gamma _{3})],\quad G=kQ/[\gamma ^{2}(x^{1}-\beta
x^{0})^{2}+(x^{2})^{2}+(x^{3})^{2}]^{3/2}.  \label{fsq}
\end{gather}%
In $S$ and in the standard basis, the basis components $F^{\mu \nu }$ of the
bivector $F$ are again obtained from (\ref{fmn}) and they are%
\begin{eqnarray}
F^{10} &=&G\gamma \lbrack (x^{1}-\beta x^{0}),\ F^{20}=G\gamma x^{2},\
F^{30}=G\gamma x^{3},  \notag \\
F^{21} &=&G\gamma \beta x^{2},\ F^{31}=G\gamma \beta x^{3},\ F^{32}=0.
\label{fi0}
\end{eqnarray}%
The expression for $F$ as a CBGQ in the $S$ frame can be found in another
way as well, i.e., to make the LT of the quantities from (\ref{q1}). Observe
that the CBGQs from (\ref{q1}) and (\ref{fsq}), which are the
representations of the bivector $F$ in $S^{\prime }$ and $S$ respectively,
are equal, $F$ from (\ref{q1}) $=$ $F$ from (\ref{fsq}); \emph{they are the
same quantity }$F$ \emph{from} (\ref{fcf}), i.e., (\ref{fmn}), \emph{for
observers in} $S^{\prime }$ \emph{and} $S$. It can be seen from (\ref{fi0})
that $F^{i0}$ \emph{and} $F^{ij}$ are different from zero for a moving
charge and they are the same as the usual components of the 3D fields $%
\mathbf{E}$ and $\mathbf{B}$, respectively. But, as already discussed and as
seen from (\ref{fc}) and (\ref{fmn}) only the whole $F$, which contains
components \emph{and the bivector basis}, is properly defined physical
quantity.\bigskip \medskip

\noindent \textit{10.2. The expressions for the 4D}\textbf{\ }$E$ \textit{and%
} $B$\bigskip

\noindent \textit{10.2.1. The general expressions\medskip }

\noindent From the known $F$ (\ref{fmn}) and\ the relations (\ref{ebv}) we
can construct in a mathematically correct way the 4D vectors $E$ and $B$ for
a charge $Q$ moving with constant velocity $u$. If written as CBGQs in the
standard basis they are given by the relation (\ref{ecb}) below

\begin{eqnarray}
E &=&E^{\mu }\gamma _{\mu }=(G/c^{2})[(u^{\nu }v_{\nu })x^{\mu }-(x^{\nu
}v_{\nu })u^{\mu }]\gamma _{\mu },  \notag \\
B &=&B^{\mu }\gamma _{\mu }=(G/c^{3})\varepsilon ^{\mu \nu \alpha \beta
}x_{\nu }u_{\alpha }v_{\beta }\gamma _{\mu },  \label{ecb}
\end{eqnarray}%
where $G$\ is from (\ref{fmn}). \emph{The vectors} $E$ \emph{and} $B$ \emph{%
are explicitly} \emph{observer dependent}, i.e., dependent on $v$. For the
same $F$ the vectors $E$ and $B$ will have different expressions depending
on the velocity of observers who measure them. It is visible from (\ref{ecb}%
) that $E$ and $B$ depend on \emph{two} velocity 4D vectors $u$ and $v$,
whereas the usual 3D vectors $\mathbf{E}$ and $\mathbf{B}$ depend only on
the 3-velocity of the charge $Q$. Note also that although $E$ and $B$ as the
CBGQs from (\ref{ecb}) depend not only on $u$ but on $v$ as well the
electromagnetic field $F$ from (\ref{fmn}) does not contain the velocity of
the observer $v$. This result directly proves that \emph{the electromagnetic
field} $F$ \emph{is the primary quantity from which the observer dependent} $%
E$ \emph{and} $B$ \emph{are derived.} The expressions for $E$ and $B$ from (%
\ref{ecb}) correctly describe fields in all cases simply specifying $u$ 
\emph{and} $v$ and this assertion holds not only for the $\left\{ \gamma
_{\mu }\right\} $ basis but for the $\{r_{\mu }\}$ basis as well, i.e., the
relation like (\ref{erc}) holds for the expressions from (\ref{ecb}).
However, observe that, as already mentioned several times, \emph{the 4D
fields} $E$ \emph{and} $B$ \emph{and the usual 3D fields} $\mathbf{E}$ \emph{%
and} $\mathbf{B}$ \emph{have the same physical interpretation only in the} $%
\gamma _{0}$ - \emph{frame} \emph{with the} $\left\{ \gamma _{\mu }\right\} $
\emph{basis in which} $E^{0}=B^{0}=0$. In Sec. 5.6.2.1 in [56] the general
expression (\ref{ecb}) for the 4D $E$ and $B$ is specified to the case when
the $\gamma _{0}$ - frame is the rest frame of the charge $Q$, the $%
S^{\prime }$ frame, $v=c\gamma _{0}^{\prime }=u$, whereas in Sec. 5.6.2.2
the same is made in the case when the $\gamma _{0}$ - frame is the
laboratory frame, the $S$ frame, $v=c\gamma _{0}$, in which the charge $Q$
is moving, $u^{\mu }=(\gamma c,\gamma \beta c,0,0)$.\bigskip

\noindent \textit{10.2.2. The} $\gamma _{0}$ \textit{- frame is the rest
frame of the charge} $Q$, \textit{the} $S^{\prime }$ \textit{frame\medskip }

\noindent If the $\gamma _{0}$ - frame is the $S^{\prime }$ frame, $%
v=c\gamma _{0}^{\prime }=u$, then (\ref{ecb}) yields that $B=0$ and only an
electric field (Coulomb field) remains, which is in agreement with the usual
3D formulation. Hence, it follows from (\ref{ecb}) that 
\begin{equation}
E=E^{\prime i}\gamma _{i}^{\prime }=Gx^{\prime i}\gamma _{i}^{\prime },\quad
E^{\prime 0}=0,\quad G=kQ/(x^{\prime i}x_{i}^{\prime })^{3/2};\quad
B=B^{\prime \mu }\gamma _{\mu }^{\prime }=0.  \label{bf}
\end{equation}%
The components in (\ref{bf}) agree, as it is expected, with the usual result
with the 3D fields, e.g., with components in Eq. (11) in [52]. Now comes the
essential difference relative to all usual approaches. In order to find the
representations of $E$ and $B$ in $S$, i.e., the CBGQs $E^{\mu }\gamma _{\mu
}$ and $B^{\mu }\gamma _{\mu }$, we can either perform the LT of $E^{\prime
\mu }\gamma _{\mu }^{\prime }$ and $B^{\prime \mu }\gamma _{\mu }^{\prime }$
that are given by (\ref{bf}), or simply to take in (\ref{ecb}) that \emph{%
both} the charge $Q$ \emph{and the \textquotedblleft
fiducial\textquotedblright\ observers} are moving relative to the observers
in $S$; $v^{\mu }=u^{\mu }=(\gamma c,\gamma \beta c,0,0)$. This yields Eq. (%
\ref{bs}) below, i.e., the CBGQs $E^{\mu }\gamma _{\mu }$ and $B^{\mu
}\gamma _{\mu }$ in $S$ with the condition that the \textquotedblleft
fiducial\textquotedblright\ observers are in $S^{\prime }$, $v=c\gamma
_{0}^{\prime }$, which is the rest frame of the charge $Q$, $u=c\gamma
_{0}^{\prime }$,

\begin{gather}
E=E^{\mu }\gamma _{\mu }=G[\beta \gamma ^{2}(x^{1}-\beta x^{0})\gamma
_{0}+\gamma ^{2}(x^{1}-\beta x^{0})\gamma _{1}+  \notag \\
x^{2}\gamma _{2}+x^{3}\gamma _{3}],\quad B=B^{\mu }\gamma _{\mu }=0,
\label{bs}
\end{gather}%
where $G$ is that one from (\ref{fsq}). The result (\ref{bs}) significantly
differs from the result obtained by the LPET, Eqs. (12a), (12b) in [52].
Under the LT, i.e., the MILT, the electric field vector transforms again to
the electric field vector and the same for the magnetic field vector. It is
worth mentioning that, in contrast to the conventional results, it holds
that $E^{\prime \mu }\gamma _{\mu }^{\prime }$ from (\ref{bf}) is $=E^{\mu
}\gamma _{\mu }$ from (75) in [56]; \emph{they are the same quantity} $E$ 
\emph{for all relatively moving inertial observers.} The same holds for $B$, 
$B^{\prime \mu }\gamma _{\mu }^{\prime }$ from (\ref{bf}) is $=B^{\mu
}\gamma _{\mu }$ from (\ref{bs}) and they are $=0$ for all observers.
Furthermore, observe that in $S^{\prime }$ there are only the spatial
components $E^{\prime i}$, whereas in $S$, as seen from (\ref{bs}), there is
also the temporal component $E^{0}$ as a consequence of the LT.\bigskip

\noindent \textit{10.2.3. The}\textbf{\ }$\gamma _{0}$ \textit{- frame is
the laboratory frame, the}\textbf{\ }$S$ \textit{frame\medskip }

\noindent Now, let us take that the \textquotedblleft
fiducial\textquotedblright\ observers are in $S$, $v=c\gamma _{0}$, in which
the charge $Q$ is moving, $u^{\mu }=(\gamma c,\gamma \beta c,0,0)$. In
contrast to the previous case, \emph{both} $E$ \emph{and} $B$ are different
from zero. The expressions for the CBGQs $E^{\mu }\gamma _{\mu }$ and $%
B^{\mu }\gamma _{\mu }$ in $S$ can be simply obtained from (\ref{ecb})
taking in it that $v=c\gamma _{0}$ and $u^{\mu }=\gamma c\gamma _{0}+\gamma
\beta c\gamma _{1}$. This yields that $E^{0}=B^{0}=0$ (from $v=c\gamma _{0}$%
) and the spatial parts are 
\begin{eqnarray}
E &=&E^{i}\gamma _{i}=G\gamma \lbrack (x^{1}-\beta x^{0})\gamma
_{1}+x^{2}\gamma _{2}+x^{3}\gamma _{3}],  \notag \\
B &=&B^{i}\gamma _{i}=(G/c)[0\gamma _{1}-\gamma \beta x^{3}\gamma
_{2}+\gamma \beta x^{2}\gamma _{3}],  \label{seb}
\end{eqnarray}%
where $G$ is again as in (\ref{fsq}). The 4D vector fields $E$ and $B$ from (%
\ref{seb}) can be compared with the usual expressions for the 3D fields $%
\mathbf{E}$ and $\mathbf{B}$ of an uniformly moving charge, e.g., from Eqs.
(12a), (12b) in [52]. It is visible that they are similar, but $E$ and $B$
in (\ref{seb}) are the 4D fields and all quantities in (\ref{seb}) are
correctly defined in the 4D spacetime, which transform by the LT, i.e., the
MILT, whereas the fields in Eqs. (12a), (12b) in [52] are the 3D fields that
transform according to the LPET.

In order to find the representations of $E$ and $B$ in $S^{\prime }$, i.e.,
the CBGQs $E^{\prime \mu }\gamma _{\mu }^{\prime }$ and $B^{\prime \mu
}\gamma _{\mu }^{\prime }$, we can either perform the LT of $E^{\mu }\gamma
_{\mu }$ and $B^{\mu }\gamma _{\mu }$ that are given by (\ref{seb}), or
simply to take in (\ref{ecb}) that relative to $S^{\prime }$ the
\textquotedblleft fiducial\textquotedblright\ observers are moving with $%
v=v^{\prime \mu }\gamma _{\mu }^{\prime }$, $v^{\prime \mu }=(c\gamma
,-\gamma \beta c,0,0)$, and the charge $Q$ is at rest relative to the
observers in $S^{\prime }$, $u^{\prime \mu }=(c,0,0,0)$. This yields the
CBGQs $E^{\prime \mu }\gamma _{\mu }^{\prime }$ and $B^{\prime \mu }\gamma
_{\mu }^{\prime }$ in $S^{\prime }$ with the condition that the
\textquotedblleft fiducial\textquotedblright\ observers are in $S$, $%
v=c\gamma _{0}$,%
\begin{eqnarray}
E &=&E^{\prime \mu }\gamma _{\mu }^{\prime }=G\gamma \lbrack -\beta
x^{\prime 1}\gamma _{0}^{\prime }+x^{\prime 1}\gamma _{1}^{\prime
}+x^{\prime 2}\gamma _{2}^{\prime }+x^{\prime 3}\gamma _{3}^{\prime }], 
\notag \\
B &=&B^{\prime \mu }\gamma _{\mu }^{\prime }=(G/c)[0\gamma _{0}^{\prime
}+0\gamma _{1}^{\prime }-\gamma \beta x^{\prime 3}\gamma _{2}^{\prime
}+\gamma \beta x^{\prime 2}\gamma _{3}^{\prime }],  \label{sc}
\end{eqnarray}%
where $G$ is as in (\ref{bf}). Again, as in the case that $v=c\gamma
_{0}^{\prime }$, it holds that $E^{\mu }\gamma _{\mu }$ from (\ref{seb}) is $%
=E^{\prime \mu }\gamma _{\mu }^{\prime }$ from (\ref{sc}); they are the same
quantity $E$ for all relatively moving inertial observers. The same holds
for $B^{\mu }\gamma _{\mu }$ from (\ref{seb}) which is $=B^{\prime \mu
}\gamma _{\mu }^{\prime }$ from (\ref{sc}) and \emph{they are both different
from zero}. Note that in this case there are only the spatial components $%
E^{i}$ in $S$, whereas in $S^{\prime }$ there is also the temporal component 
$E^{\prime 0}$ as a consequence of the MILT.

It is visible from (\ref{sc}) that if the $\gamma _{0}$ - frame is the lab
frame ($v=c\gamma _{0}$) in which the charge $Q$ is moving then $E^{\prime
\mu }\gamma _{\mu }^{\prime }$ and $B^{\prime \mu }\gamma _{\mu }^{\prime }$
in the rest frame of the charge $Q$, the $S^{\prime }$ frame, are completely
different than those from (\ref{bf}); in (\ref{sc}) $B^{\prime \mu }\gamma
_{\mu }^{\prime }$ is different from zero and the representation of $E$\
contains also the term $E^{\prime 0}\gamma _{0}^{\prime }$.

It has to be emphasized that \emph{all four expressions for} $E$ \emph{and} $%
B$, (\ref{bf}), (\ref{bs}), (\ref{seb}) \emph{and} (\ref{sc}), \emph{are the
special cases of} $E$ \emph{and} $B$ \emph{given by} (\ref{ecb}). \emph{They
all give the same} $F$ \emph{from} (\ref{fmn}), \emph{which is the
representation }(\emph{CBGQ})\emph{\ of} $F$ \emph{given by the basis free,
abstract, bivector} (\ref{fcf}).\bigskip \bigskip

\noindent \textbf{11. Comparison with experiments}\bigskip

\noindent It is usually considered that the LPET of the 3D $\mathbf{E}$ and $%
\mathbf{B}$, Eq. (11.149) in [2], i.e., (\ref{JCB}) here, are firmly
confirmed by experiments and, accordingly, that there are not separate
electric and magnetic 4D vectors. In particular, it is considered that in
the rest frame of the charge the 3D electric field is given, e.g., by Eq.
(11) in [52], the 3D magnetic field is zero, whereas in a relatively moving
inertial frame the 3D vectors are given by Eqs. (12a), (12b) in [52] and
they are obtained, as stated in [52], by the LT of the electric field of the
point charge from $S^{\prime }$\ to $S$\ frame. In [52], these LT, are what
we call the LPET of the 3D vectors, Eq. (11.149) in [2], i.e., (\ref{JCB})
here. However, note that the 3D fields (11) in [52] for a charge at rest and
(12a), (12b) in [52] for an uniformly moving charge are usually obtained as
the solutions of Maxwell's equations without the use of the LT. In that case
both 3D fields are determined in the same frame, usually it is the
laboratory frame, but they refer to a charge at rest in that frame and to an
uniformly moving charge in the same frame. All experiments are made \emph{%
only} in the laboratory frame in which the fields are measured for the two
mentioned states of motion. The observers are in both cases \emph{only} in
the laboratory frame. Hence it is not true that the LPET of the 3D $\mathbf{E%
}$ and $\mathbf{B}$ fields are firmly confirmed by experiments. As seen from
the preceding sections if the observers are at rest in the laboratory frame
and they use the standard basis, i.e., the lab frame is the $\gamma _{0}$ -
frame, then $E^{0}=B^{0}=0$ and the spatial components of the 4D $E$ and $B$
agree with the components of the 3D $\mathbf{E}$ and $\mathbf{B}$ for both,
a charge at rest in the lab frame, equation (\ref{bf}) and an uniformly
moving charge in the lab frame, equation (\ref{seb}). This shows that in the
cases in which only the fields are investigated the 4D fields are in the
same agreement with \emph{existing} experiments as are the 3D fields. From
that result one could think that the 3D fields can explain all experiments
and that there is no need for the 4D fields.

However, as shown previously, the formulation with the 4D GQs is in a true
agreement, independent of the chosen inertial reference frame and of the
chosen system of coordinates in it, with experiments in electromagnetism,
the motional emf in [18, 31], the Faraday disk in [19] and the Trouton-Noble
experiment in [33, 55]. \emph{As shown in the mentioned papers} [18, 19, 31,
33, 55] \emph{it is not the case with the usual 3D formulation}.

Thus, for example, in section 5.1. in [18] the motional emf $\varepsilon $
is calculated using the 3D quantities (the Lorentz force as a 3D vector, $%
\mathbf{F}_{L}\mathbf{=}q\mathbf{E}+q\mathbf{u}\times \mathbf{B}$, and $%
\varepsilon =\oint (\mathbf{F}_{L}\mathbf{/}q)\cdot \mathbf{dl}$, Eq. (26)
in [18]) and their LPET, Eq. (11.149) in [2], i.e., (\ref{JCB}) here. In
section 5.2 in [18] $\varepsilon $ is calculated using the 4D GQs and their
mathematically correct LT, i.e., the MILT, like (\ref{LTE}). The Lorentz
force $K_{L}$ is defined as in equations (\ref{lfa}) or (\ref{lf}). \emph{%
The emf} $\varepsilon $ \emph{is defined as an invariant 4D quantity, the
Lorentz scalar}, Eq. (35) in [18], $\varepsilon =\int_{\Gamma
}(K_{L}/q)\cdot dl=\int_{\Gamma }(K_{L}^{\mu }/q)dl_{\mu }=(1/c)\int_{\Gamma
}F^{\mu \nu }u_{\nu }dl_{\mu }$, where \emph{vector} $dl$ is the
infinitesimal spacetime length and $\Gamma $ is the spacetime curve. In
section 5.1 in [18] it is shown that the emf obtained by the application of
the LPET is different for relatively moving 4D observers, $\varepsilon =VBl$
in $S$ (the laboratory frame) and $\varepsilon ^{\prime }=\gamma VBl$ in $%
S^{\prime }$,\ Eqs. (27) and (29) respectively, which means that \emph{the
principle of relativity is not satisfied in the usual formulation of
electromagnetism with the 3D quantities and their LPET} of $\mathbf{E}$ and $%
\mathbf{B}$. On the other hand, if the 4D GQs and their MILT, like (\ref{LTE}%
), are used then the emf is always the same; it is independent of the chosen
reference frame and of the chosen system of coordinates in it. Thus, if $%
\varepsilon $ is defined as an invariant 4D quantity, the Lorentz scalar,
Eq. (35) in [18], then always the same value for $\varepsilon $ is obtained, 
$\varepsilon =\gamma VBl$, Eqs. (36) and (37) in [18]. These results
unambiguously show that \emph{the principle of relativity is naturally
satisfied in the approach with 4D GQs and their mathematically correct MILT}%
, like (\ref{LTE}). The result that the conventional theory with the 3D $%
\mathbf{E}$ and $\mathbf{B}$ and their LPET, Eqs. (11.148) and (11.149) in
[2], i.e., (\ref{JCB}) here, yields different values for the motional emf $%
\varepsilon $ for relatively moving inertial observers, whereas the approach
with 4D GQs and their MILT yields always the same value for $\varepsilon $, $%
\varepsilon =\gamma VBl$, is very strong evidence that the approach with 4D
GQs is a relativistically correct approach. \emph{It is for the
experimentalists to find the way to measure the emf} $\varepsilon $ \emph{%
with a great precision in order to see that in the laboratory frame} $%
\varepsilon =\gamma VBl$ \emph{and not simply} $\varepsilon =VBl$.

Such an experiment would be a crucial experiment that could verify from the
experimental viewpoint the validity of the formulation of the
electromagnetism with the 4D GQs and their mathematically correct LT, MILT, (%
\ref{aec}) - (\ref{el}). The same result as in Sec. 5.2 in [18] is obtained
in [31] but exclusively dealing with $F$ and not with its decompositions (%
\ref{E2}) and (\ref{fm}). As mentioned above the same difference between the
usual approach with 3D fields and the approach with 4D GQs is shown to exist
in the case of the Faraday disk in [19].

Furthermore, as already mentioned in Sec. 9. in the approach with 4D GQs and
their MILT, like (\ref{LTE}) (the Trouton-Noble paradox [33, 55], Jackson's
paradox [48, 49] and the \textquotedblleft charge-magnet
paradox\textquotedblright\ [56]) there is no paradox \ and thus there is no
need for some resolutions of the paradoxes and there is no need for the
introduction of some \textquotedblleft hidden\textquotedblright\ quantities
[57-61] or the Einstein-Laub force [52-54].

As already stated, in [32], the constitutive relations and the
magnetoelectric effect in moving media are explained in a completely new way
using 4D GQs and their mathematically correct MILT. In equation (17) in [32]
it is shown how the polarization vector $P(x)$ depends on $E$, $B$, $u$, the
bulk velocity vector of the medium and $v$, the velocity vector of the
observer who measures fields, $P^{\mu }\gamma _{\mu }=(\varepsilon _{0}\chi
_{E}/c)[(1/c)(E^{\mu }v^{\nu }-E^{\nu }v^{\mu })+\varepsilon ^{\mu \nu
\alpha \beta }v_{\alpha }B_{\beta }]u_{\nu }\gamma _{\mu }$, whereas in
equation (18) in [32] the same is shown for the magnetization vector $M(x)$, 
$M^{\mu }\gamma _{\mu }=\varepsilon _{0}\chi _{B}[(B^{\mu }v^{\nu }-B^{\nu
}v^{\mu })+(1/c)\varepsilon ^{\mu \nu \alpha \beta }E_{\alpha }v_{\beta
}]u_{\nu }\gamma _{\mu }$. The last term in the expression for $P^{\mu
}\gamma _{\mu }$\ and the last term in the expression for $M^{\mu }\gamma
_{\mu }$\ describe the magnetoelectric effect in a moving dielectric.
According to the last term in $P^{\mu }\gamma _{\mu }$ a moving dielectric
becomes electrically polarized if it is placed in a magnetic field, the
Wilsons' experiment, reference [29] in [32]. Similarly, the last term in $%
M^{\mu }\gamma _{\mu }$ shows that a moving dielectric becomes magnetized if
it is placed in an electric field, R\"{o}ntgen's experiment, reference [30]
in [32].\bigskip \bigskip

\noindent \textbf{12. Discussion and conclusion}\bigskip

\noindent The main point in the whole paper is that in the 4D spacetime
physical laws are geometric, coordinate-free relationships between the 4D
geometric, coordinate-free quantities. This point of view is also adopted in
the nice textbook [24] (and, as well, in [47]) but not in the consistent
way. They still introduce the 3D vectors and their transformations, e.g., in
Sec. 1.10 in [24] and this is discussed in Sec. 8. here. A fully consistent
application of this viewpoint is adopted in Oziewicz's papers, see, e.g.,
[28]. The same viewpoint is adopted in all my papers given in the
references, including the present paper. Particularly, in [62], under the
title \textquotedblleft Nature of Electric and Magnetic Fields; How the
Fields Transform\textquotedblright\ we have already presented many results
that are given in this paper. Here, in this paper, \emph{the mathematically
correct proofs} are given that in this geometric approach, i.e., in the ISR,
the electric and magnetic fields are properly defined vectors on the 4D
spacetime, Secs. 3.1. and 3.3. According to Oziewicz's proof from Sec. 3.1.,
e.g., the electric field vector must have four components (some of them can
be zero) since it is defined on the 4D spacetime and not, as usually
considered, only three components. In Sec. 3.3. it is taken into account
that, as proved in [33], the primary quantity for the whole electromagnetism
is the electromagnetic field bivector $F$.\ The decomposition of $F$\ given
by Eq. (\ref{E2}) expresses $F$\ in terms of observer dependent electric and
magnetic 4D vectors $E$ and $B$, which are given by Eq. (\ref{E1}).\ Both,
Eqs. (\ref{E2}) and (\ref{E1}), are with the abstract, coordinate-free
quantities. This is in a sharp contrast with the usual covariant approaches,
e.g., [2, 25] in which it is considered that $F^{\alpha \beta }$ (the
components implicitly taken in the standard basis) is physically
well-defined quantity. Moreover, these components are considered to be six
indepent components of the 3D $\mathbf{E}$ and $\mathbf{B}$, see Eqs. (\ref%
{ieb}) and (\ref{eb2}). Then, as described in Sec. 7., in these approaches
[2, 25],\ \emph{the transformations of the components of} $\mathbf{E}$ \emph{%
and} $\mathbf{B}$ (\ref{ee}) \emph{are obtained supposing that they
transform under the LT as the components of} $F^{\alpha \beta }$ \emph{%
transform}, Eqs. (\ref{fe}) and (\ref{ee}). The objections to such treatment
are given in Sec. 7., the objections 1) - 5). From the mathematical
viewpoint all these objections are well-founded since they are based on the
following facts:

1) The bivector $F(x)$, as described in detail in [33] and very briefly in
Sec. 3.2. here, is determined, for the given sources, by the solutions of
the equation (\ref{fef}), i.e., (\ref{mf}) (with CBGQs in the $\left\{
\gamma _{\mu }\right\} $ basis) and not by the components of the 3D $\mathbf{%
E}$ and $\mathbf{B}$. It is a 4D GQ and not only components. It yields a
complete description of the electromagnetic field without the need for the
introduction either the field vectors or the potentials.

2) As seen from Sec. 2. and particularly from Eqs. (\ref{are}) and (\ref{FEr}%
) the identification of the components of the 3D $\mathbf{E}$ and $\mathbf{B}
$ with the components of $F^{\alpha \beta }$ is synchronization dependent.
Moreover, it is completely meaningless in the \textquotedblleft
r\textquotedblright\ synchronization, i.e., in the $\left\{ r_{\mu }\right\} 
$ basis, that is discussed and explained in Sec. 2. Both bases, the commonly
used standard basis with Einstein's synchronization and the $\left\{ r_{\mu
}\right\} $ basis with the \textquotedblleft r\textquotedblright\
synchronization are equally well physical and relativistically correct
bases. It is worth mentioning that in [63], in which a geometric approach
with exterior forms is used, it is considered that the usual identification,
(\ref{ieb}), is premetric, but, as explained above, it is synchronization
dependent and thus dependent on the chosen metric of the 4D spacetime. This
is discussed in detail in Sec. 5.3 in [32] in connection with the
constitutive relations.

Furthermore, it is proved in Sec. 4.1. with the coordinate-free quantities
and the active LT and in Sec. 4.2. with CBGQs and the passive LT that the
mathematically correct LT, the MILT, of, e.g., the electric field vector are
given by (\ref{aec}) - (\ref{el}) and not by the LPET of the 3D vectors Eqs.
(11.148) and (11.149) in [2], i.e., Eq. (\ref{ee}) or Eq. (\ref{ut}) here.

In Sec. 5.1. the same fundamental difference between the correct LT, the
MILT, and the LPET of the 3D vectors is explicitly exposed using matrices.
The equations (\ref{ecm}) - (\ref{an}) refer to the correct LT, the MILT, of
the components in the standard basis of the electric field 4D vector in
which the transformed components $E^{\prime \mu }$\ are obtained as $%
E^{\prime \mu }=c^{-1}F^{\prime \mu \nu }v_{\nu }^{\prime }$, i.e., \emph{%
both} $F^{\mu \nu }$\emph{\ and the velocity of the observer }$v=c\gamma
_{0} $\emph{\ are transformed}$\ $by the matrix of the LT $A_{\nu }^{\mu }$
(the boost in the direction $x^{1}$). It is visible from Eq. (\ref{an}) that
the same components are obtained as $E^{\prime \mu }=A_{\nu }^{\mu }E^{\nu }$%
\ and they are the same as in (\ref{LTE}). This means that \emph{under the
mathematically correct LT, the MILT, the electric field 4D vector transforms
again only to the electric field 4D vector as any other 4D vector transforms.%
} As stated at the end of Sec. 5.1. if $E$\ is written as a CBGQ then again
holds the relation (\ref{erc}) as for any other CBGQ. On the other hand Eq. (%
\ref{Em4}) refers to the LPET in which the transformed components $%
E_{F}^{\prime \mu }$\ are obtained as $E_{F}^{\prime \mu }=c^{-1}F^{\prime
\mu \nu }v_{\nu }$, i.e., only $F^{\mu \nu }$ \emph{is transformed by the LT
but not the velocity of the observer} $v=c\gamma _{0}$. These transformed
components $E_{F}^{\prime \mu }$\ are the same as in Eq. (\ref{ut}). The
transformed spatial components $E_{F}^{\prime i}$\ are the same as are the
transformed components of the usual 3D vector $\mathbf{E}$, i.e., as in Eq.
(11.148) in [2].\ However, according to these transformations the 4D vector
with $E^{0}=0$ is transformed in such a way that the transformed temporal
component is again zero, $E_{F}^{\prime 0}=0$. Hence, as stated in Sec.
5.1., such transformations cannot be the mathematically correct LT.

It can be concluded from the whole consideration in this paper that in the
4D spacetime an independent physical reality has to be attributed to the 4D
geometric quantities, coordinate-free quantities or the CBGQs, e.g., the
electromagnetic field bivector $F$, the 4D vectors of the electric $E$\ and
magnetic $B$ fields, etc., and not to the usual 3D quantities, e. g., the 3D 
$\mathbf{E}$ and $\mathbf{B}$. This is the answer to the question what is
the nature of the electric and magnetic fields. Furthermore, the
mathematically correct LT are properly defined on the 4D spacetime. They can
correctly transform only the 4D quantities like $E$\ and $B$, the
mathematically correct LT, the MILT, (\ref{aec}) - (\ref{el}), according to
which, e.g., \emph{the electric field 4D vector transforms again only to the
electric field 4D vector as any other 4D vector transforms.} The LT cannot
act on the 3D quantities like the 3D $\mathbf{E}$ and $\mathbf{B}$, which
means that the LPET of the 3D quantities, e.g., the 3D vectors $\mathbf{E}$
and $\mathbf{B}$, Eqs. (11.148) and (11.149) in [2], i.e., Eq. (\ref{ee}) or
Eq. (\ref{ut}) here, are not properly defined LT in the 4D spacetime. This
is the answer to the question how the fields transform.

Here, it is worth mentioning that in [26] a fundamental result is obtained
by a consistent application of the 4D GQs and the relations like (\ref{fm})
and (\ref{ebv}). First, the generalized Uhlenbeck-Goudsmit hypothesis is
formulated as the relation which connects the dipole moment tensor $D^{ab}$
and the spin 4D tensor $S^{ab}$, $D^{ab}=g_{S}S^{ab}$, Eq. (9) in [26],
instead of the usual relation between the 3D vectors, the magnetic moment $%
\mathbf{m}$ and the spin 3D vector $\mathbf{S}$, $\mathbf{m}=\gamma _{S}%
\mathbf{S}$. Then, both $D^{ab}$ and $S^{ab}$ are decomposed like in (\ref%
{fm}) into the dipole moment 4D vectors $m^{a}$, $d^{a}$, Eq. (2) in [26],
and the\ intrinsic angular momentum 4D vectors, the usual $S^{a}$ and the
new one $Z^{a}$, Eq. (8) in [26]. It is obtained in a mathematically correct
procedure that $d^{a}$, the electric dipole moment of a fundamental
particle, is determined by $Z^{a}$ and not, as generally accepted, by the
spin 3D vector $\mathbf{S}$. Observe that in [26] an abstract index notation
is used, $m^{a}$, $d^{a}$, $S^{ab}$, ... are the 4D geometric,
coordinate-free quantities. \bigskip \bigskip

\noindent \textbf{Acknowledgments\bigskip }

\noindent I am cordially thankful to late Zbigniew Oziewicz for numerous and
very useful discussions during years and for the continuos support of my
work. It is a pleasure to acknowledge to Larry Horwitz and Martin Land for
inviting me to the IARD conferences, for the valuable discussions and for
the continuos support of my work. I am also grateful to Alex Gersten for
useful discussions and for the continuos support of my work.\bigskip \bigskip

\noindent \textbf{Appendix\bigskip }

\noindent In this Appendix we briefly describe the essential differences
between the 4D geometric approach, the ISR, and Einstein's definition of the
Lorentz contraction, e.g., for a moving rod. This is explained in detail in
Secs. 2. - 2.3. in [8] and Secs. 3.1., 4.1. and Figs. 1. and 3. in [6].
Here, the mathematical formalism is different than in [8] and [6]. In the
geometric approach one deals with the abstract 4D geometric quantities,
i.e., with the position vectors $x_{A}$, $x_{B},$ of the events $A$ and $B$,
respectively, with the distance vector $l_{AB}=x_{B}-x_{A}$ and with the
spacetime length, $l=L_{0}$, see (\ref{sl}). The essential feature of the
geometric approach is that \emph{any abstract 4D geometric quantity}, e.g.,
the distance vector $l_{AB}=x_{B}-x_{A}$, \emph{is} \emph{only one quantity,
the same quantity in the 4D spacetime} for all relatively moving frames of
reference and for all systems of coordinates that are chosen in them. The
abstract vector $l_{AB}$ can be decomposed in different bases and then these
representations, the CBGQs, of the same abstract 4D geometric quantity $%
l_{AB}$ contain both the basis components and the basis vectors. Let us
explain it taking a particular choice for $l_{AB}$, which in the usual
\textquotedblleft 3+1\textquotedblright\ picture corresponds to a rod that
is at rest in an inertial frame of reference (IFR) $S$ (with the standard
basis in it) and situated along the common $x^{1}$, $x^{\prime 1}$ $-$ axes.
Its rest length is denoted as $L_{0}$. The situation is depicted in Fig. 1.
in [6]. $l_{AB}$ is decomposed, i.e., it is written as a CBGQ, in the
standard basis and in $S$ and $S^{\prime }$, where the rod is moving, as 
\begin{equation}
l_{AB}=l_{AB}^{\mu }\gamma _{\mu }=0\gamma _{0}+L_{0}\gamma
_{1}=l_{AB}^{\prime \mu }\gamma _{\mu }^{\prime }=-\beta \gamma L_{0}\gamma
_{0}^{\prime }+\gamma L_{0}\gamma _{1}^{\prime },  \label{dv}
\end{equation}%
As already stated several times, the components $l_{AB}^{\mu }$ are
transformed by the LT and the basis vectors $\gamma _{\mu }$ by the inverse
LT leaving the whole CBGQ unchanged. In $S$, the position vectors $x_{A,B}$
are determined simultaneously, $x_{B}^{0}-x_{A}^{0}=l_{AB}^{0}=0$, i.e., the
temporal part of $l_{AB}^{\mu }$ is zero. In the standard basis, which is
commonly used in the usual approaches, there is a dilation of the spatial
part $l_{AB}^{\prime 1}=\gamma L_{0}$ with respect to $l_{AB}^{1}=L_{0}$ and
not the Lorentz contraction as predicted in Einstein's formulation of SR.
Similarly, as explicitly shown in [8] and [6], in the $\{r_{\mu }\}$ basis,
i.e., with the \textquotedblleft r\textquotedblright\ synchronization, if
only spatial parts of $l_{AB,r}^{\mu }$ and $l_{AB,r}^{\prime \mu }$ are
compared then one finds the dilation $\infty \succ l_{AB,r}^{\prime 1}\geq
L_{0}$ for all $\beta _{r}$. However, the comparison of only spatial parts
of the components of the distance vector $l_{AB}$ in $S$ and $S^{\prime }$
is physically meaningless in the geometric approach,\ since \emph{some
components of the tensor quantity, when they are taken alone, do not
correspond to some definite 4D physical quantity.} Note that if $%
l_{AB}^{0}=0 $ then the LT yield that $l_{AB}^{\prime \mu }$ in any other
IFR $S^{\prime } $ contains the time component as well, $l_{AB}^{\prime
0}=x_{B}^{\prime 0}-x_{A}^{\prime 0}=-\beta \gamma L_{0}\neq 0$. Hence, 
\emph{the LT yield that the spatial ends of the rod are not determined
simultaneously in} $S^{\prime }$, \emph{i.e., the temporal part of} $%
l_{AB}^{\prime \mu }$ \emph{is not zero.} For the spacetime length $l$ it
holds that%
\begin{equation}
l^{2}=\mid l_{AB}^{\mu }l_{AB,\mu }\mid =\mid l_{AB}^{\prime \mu }l_{AB,\mu
}^{\prime }\mid =\mid l_{AB,r}^{\mu }l_{AB,r,\mu }\mid =L_{0}^{2}.
\label{sl}
\end{equation}%
In $S$, the rest frame of the rod, where the temporal part of $l_{AB}^{\mu }$
is $l_{AB}^{0}=0,$ the spacetime length $l$ is a measure of the spatial
distance, i.e., of the rest spatial length of the rod, as in the
prerelativistic physics. \emph{The observers in all other IFRs will
\textquotedblleft look\textquotedblright\ at the same events} $A$ \emph{and} 
$B$, \emph{the same distance vector} $l_{AB}$ \emph{and the same spacetime
length} $l$, \emph{but associating with them different coordinates; it is
the essence of the geometric approach. They all obtain the same value} $l$ 
\emph{for the spacetime length,} $l=L_{0}$.

It is worth mentioning, once again, that the 4D geometric treatment with $%
l_{AB}$ and $l$ is a generalization and a mathematically better founded
formulation of the ideas expressed by Rohrlich [11] and Gamba [64]. Indeed,
Rohrlich [11] states: \textquotedblright A quantity is therefore physically
meaningful (in the sense that it is of the same nature to all observers) if
it has tensorial properties under Lorentz
transformations.\textquotedblright\ Similarly Gamba [64], when discussing
the sameness of a physical quantity (for example, \emph{a nonlocal quantity} 
$A_{\mu }(x_{\lambda },X_{\lambda })$, which is a function of two points in
the 4D spacetime $x_{\lambda }$ and $X_{\lambda }$) for different inertial
frames of reference $S$ and $S^{\prime }$, declares: \textquotedblright The
quantity $A_{\mu }(x_{\lambda },X_{\lambda })$ for $S$ is the same as the
quantity $A_{\mu }^{\prime }(x_{\lambda }^{\prime },X_{\lambda }^{\prime })$
for $S^{\prime }$ when all the primed quantities are obtained from the
corresponding unprimed quantities through Lorentz transformations (tensor
calculus).\textquotedblright\ Rohrlich and Gamba worked with the usual
covariant approach, i.e., with the components implicitly taken in the
standard basis, which means that only Einstein's synchronization is
considered to be physically admissible. The quantities $A_{\mu }(x_{\lambda
},X_{\lambda })$ and $A_{\mu }^{\prime }(x_{\lambda }^{\prime },X_{\lambda
}^{\prime })$ refer to the same physical quantity, but they are not
mathematically equal quantities since bases are not included. In the
approach with the 4D geometric quantities, i.e., in the ISR, one deals with
mathematically equal quantities, e.g., for a nonlocal quantity $%
l_{AB}=x_{B}-x_{A}$ it holds that

\begin{equation}
l_{AB}=l_{AB}^{\mu }\gamma _{\mu }=l_{AB}^{\prime \mu }\gamma _{\mu
}^{\prime }=l_{AB,r}^{\mu }r_{\mu }=l_{AB,r}^{\prime \mu }r_{\mu }^{\prime
}=..,  \label{dv1}
\end{equation}%
where the primed quantities are the Lorentz transforms of the unprimed ones.
In order to treat different systems of coordinates on an equal footing we
have derived a form of the LT that is independent of the chosen system of
coordinates, including different synchronizations, see Eq. (2) in [8], or
Eq. (1) in [6]. Also, Eq. (4) in [6], it is presented the transformation
matrix that connects Einstein's system of coordinates with another system of
coordinates in the same reference frame.

On the other hand, as shown in Sec. 2.2. in [8] and Sec. 4.1. and Fig. 3. in
[6], in Einstein's formulation of SR, instead of to work with geometric
quantities $x_{A,B}$, $l_{AB}$ and $l$ one deals only with the spatial, or
temporal, \emph{components} of their coordinate representations $x_{A}^{\mu
} $, $x_{B}^{\mu }$ and $l_{AB}^{\mu }$ \emph{in the standard basis}. The
geometric character of physical quantities, i.e., the basis vectors, and
some asymmetric synchronization, e.g., the \textquotedblleft
r\textquotedblright\ synchronization, which is equally physical as the
Einstein synchronization, are never taken into account. According to
Einstein's definition [1] of the spatial length the spatial ends of the rod
must be taken simultaneously for the observer, i.e., \emph{he} \emph{defines
length as the spatial distance between two spatial points on the (moving)
object measured by simultaneity in the rest frame of the observer.} In the
4D (here, for simplicity, as in [8] and [6], we deal only with 2D) spacetime
and in the $\{\gamma _{\mu }\}$ basis the simultaneous events $A$ and $B$
(whose spatial parts correspond to the spatial ends of the rod) are the
intersections of $x^{1}$ axis (that is along the spatial basis vector $%
\gamma _{1}$) and the world lines of the spatial ends of the rod that is at
rest in $S$ and situated along the $x^{1}$ axis. The components of the
distance vector are $l_{AB}^{\mu }=x_{B}^{\mu }-x_{A}^{\mu }=(0,L_{0})$; for
simplicity, it is taken that $t_{B}=t_{A}=a=0$. Then in $S$, the rest frame
of the object, the spatial part $l_{AB}^{1}=L_{0}$ of $l_{AB}^{\mu }$ is
considered to define the rest spatial length. Furthermore, one uses the
inverse LT to express $x_{A}^{\mu }$, $x_{B}^{\mu }$ and $l_{AB}^{\mu }$ in $%
S$ in terms of the corresponding quantities in $S^{\prime }$, in which the
rod is moving. This procedure yields 
\begin{eqnarray}
l_{AB}^{0} &=&ct_{B}-ct_{A}=\gamma (l_{AB}^{\prime 0}+\beta l_{AB}^{\prime
1}),  \notag \\
l_{AB}^{1} &=&x_{B}^{1}-x_{A}^{1}=\gamma (l_{AB}^{\prime 1}+\beta
l_{AB}^{\prime 0}).  \label{lc}
\end{eqnarray}%
Now, instead of to work with 4D tensor quantities and their LT, as in the 4D
geometric approach, in the usual formulation one forgets about the
transformation of the temporal part $l_{AB}^{0}$, the first equation in (\ref%
{lc}), and considers only the transformation of the spatial part $l_{AB}^{1}$%
, the second equation in (\ref{lc}). Furthermore, \emph{in that relation for}
$l_{AB}^{1}$ \emph{one assumes that} $t_{B}^{\prime }=t_{A}^{\prime
}=t^{\prime }=b$, i.e., \emph{that} $x_{B}^{\prime 1}$ \emph{and} $%
x_{A}^{\prime 1}$ \emph{are simultaneously determined at some arbitrary} $%
t^{\prime }=b$ in $S^{\prime }$. However, \emph{in 4D} (at us 2D) spacetime
such an assumption means that \emph{in} $S^{\prime }$ \emph{one does not
consider the same events} $A$ \emph{and} $B$ \emph{as in} $S$ but some other
two events $C$ and $D$, which means that $t_{B}^{\prime }=t_{A}^{\prime }$
has to be replaced with $t_{D}^{\prime }=t_{C}^{\prime }=b$. The events $C$
and $D$ are the intersections of the line (the hypersurface $t^{\prime }=b$
with arbitrary $b$) parallel to the spatial axis $x^{\prime 1}$ (which is
along the spatial base vector $\gamma _{1}^{\prime }$) and of the above
mentioned world lines of the spatial end points of the rod. Then, in the
above transformation for $l_{AB}^{1}$ (\ref{lc}) one has to write $%
x_{D}^{\prime 1}-x_{C}^{\prime 1}=l_{CD}^{\prime 1}$ instead of $%
x_{B}^{\prime 1}-x_{A}^{\prime 1}=l_{AB}^{\prime 1}$. The spatial parts $%
l_{AB}^{1}$ and $l_{CD}^{\prime 1}$ are the \emph{spatial distances }between
the events $A$, $B$ and $C$, $D$, respectively. \emph{In Einstein's
formulation, the spatial distance}\textit{\ }$%
l_{AB}^{1}=x_{B}^{1}-x_{A}^{1}=L_{0}$ \emph{defines the spatial length of
the rod at rest in }$S$,\emph{\ whereas} $l_{CD}^{\prime 1}=x_{D}^{\prime
1}-x_{C}^{\prime 1}$ \emph{is considered to define the spatial length of the
moving rod in}\textit{\ }$S^{\prime }$. Hence, from the equation for $%
l_{AB}^{1}$ (\ref{lc}) one finds the relation between $l^{\prime
1}=l_{CD}^{\prime 1}$ and $l^{1}=l_{AB}^{1}=L_{0}$ as the famous formula for
the Lorentz contraction of the moving rod 
\begin{equation}
l^{\prime 1}=x_{D}^{\prime 1}-x_{C}^{\prime 1}=L_{0}/\gamma
=(x_{B}^{1}-x_{A}^{1})/\gamma ,\,\,\,\mathrm{with}\,\,t_{C}^{\prime
}=t_{D}^{\prime },\,\,\,\mathrm{and}\,\,\,t_{B}=t_{A},  \label{contr}
\end{equation}%
where $\gamma =(1-\beta ^{2})^{-1/2}$, $\beta =U/c$ and $U=\left\vert 
\mathbf{U}\right\vert $; $\mathbf{U}$ is the 3-velocity of $S^{\prime }$
relative to $S$. As can be nicely seen from Fig. 3 in [6], the spatial
lengths $L_{0}$ and $l_{CD}^{\prime 1}$ refer not to the same 4D tensor
quantity,\ as in the 4D geometric approach, see Fig. 1 in [6], but to two
different quantities, \emph{two different set of events in the 4D spacetime}%
. These quantities are obtained by the same measurements in $S$ and $%
S^{\prime }$; the spatial ends of the rod are measured simultaneously at
some $t=a$ in $S$ and also at some $t^{\prime }=b$ in $S^{\prime }$. But $a$
in $S$ and $b$ in $S^{\prime }$ are not related by the LT or any other
coordinate transformation. This means that the Lorentz contraction, as
already shown by Rohrlich [11] and Gamba [64], is a typical example of an
\textquotedblleft apparent\textquotedblright\ transformation. \emph{It has
nothing in common with the LT of the 4D geometric quantities.} We see that
in Einstein's approach [1] the spatial and temporal parts of events are
treated separately, and moreover the time component is not transformed in
the transformation that is called - the Lorentz contraction. In addition, as
can be seen from Sec. 4.1. and Fig. 3 in [6], in Einstein's approach [1] the
considered effect is dependent on the chosen synchronization. If the
\textquotedblleft r\textquotedblright\ synchronization is used, then there
is not only the usual Lorentz contraction of the moving rod but also a
length dilation depending on $\beta _{r}$. Thus, contrary to the generally
accepted opinion, \emph{the Lorentz contraction is not a well-defined
relativistic effect in the 4D spacetime}. As seen from Fig. 4 in [6] the
similar conclusion holds for the usual time dilation of the moving clock.
The relativistically, i.e., mathematically, correct treatments of a moving
rod and a moving clock are presented in Figs. 1 and 2 in [6].\bigskip
\bigskip

\noindent \textbf{References\bigskip }

\noindent \lbrack 1] A. Einstein, Ann. Phys. \textbf{17,} 891 (1905);
translated by W. Perrett and

G. B. Jeffery in: \textit{The Principle of Relativity }(Dover, New York,
1952).

\noindent \lbrack 2] J. D. Jackson, \textit{Classical Electrodynamics}
(Wiley, New York, 1998), 3rd ed.

\noindent \lbrack 3] H. A. Lorentz, Proceedings of the Royal Netherlands
Academy of Arts and Sciences \textbf{6,} 809 (1904).

\noindent \lbrack 4] H. Poincar\'{e}, Rend. del Circ. Mat. di Palermo 
\textbf{21,} 129 (1906).

\noindent \lbrack 5] A. A. Logunov, Hadronic J. \textbf{19,} 109 (1996).

\noindent \lbrack 6] T. Ivezi\'{c}, Found. Phys. \textbf{31,} 1139 (2001).

\noindent \lbrack 7] T. Ivezi\'{c}, Found. Phys. Lett. \textbf{15,} 27
(2002); arXiv: physics/0103026 (2001); arXiv: physics/0101091 (2001).

\noindent \lbrack 8] T. Ivezi\'{c}, Found. Phys. Lett. \textbf{12,} 507
(1999); arXiv: physics/0102014 (2012).

\noindent \lbrack 9] T. Ivezi\'{c}, Ann. Fond. Louis de Broglie \textbf{27},
287 (2002).

\noindent \lbrack 10] T. Ivezi\'{c}, J. Phys.: Conf. Ser. \textbf{437},
012014 (2013); arXiv: 1204.5137 (2012).

\noindent \lbrack 11] F. Rohrlich, Nuovo Cimento B \textbf{45}, 76 (1966).

\noindent \lbrack 12] D. Hestenes, \textit{Space-Time Algebra (}Gordon \&
Breach, New York, 1966)

\noindent \lbrack 13] Am. J Phys. \textbf{71,} 691 (2003)

\noindent \lbrack 14] D. Hestenes and G. Sobczyk, \textit{Clifford Algebra
to Geometric Calculus }(Reidel, Dordrecht, 1984)

\noindent \lbrack 15] C. Doran and A. Lasenby, \textit{Geometric algebra for
physicists }(Cambridge University Press, Cambridge, 2003).

\noindent \lbrack 16] H. Minkowski, Nachr. Ges. Wiss. G\"{o}ttingen 53
(1908); reprinted in: Math. Ann. \textbf{68,} 472 (1910); English
translation in: M. N. Saha and S. N. Bose, \textit{The Principle of
Relativity: Original Papers by A. Einstein and H. Minkowski }(Calcutta
University Press, Calcutta, 1920).

\noindent \lbrack 17] T. Ivezi\'{c}, Found. Phys. \textbf{33,} 1339 (2003).

\noindent \lbrack 18] T. Ivezi\'{c}, Found. Phys. Lett. \textbf{18, }301
(2005).

\noindent \lbrack 19] T. Ivezi\'{c}, Found. Phys. \textbf{35,} 1585 (2005).

\noindent \lbrack 20] T. Ivezi\'{c}, Fizika A\textit{\ }\textbf{17,} 1
(2008); arXiv: physics/0607189 (2006).

\noindent \lbrack 21] T. Ivezi\'{c}, arXiv:\textit{\ }0809.5277 (2008).

\noindent \lbrack 22] T. Ivezi\'{c}, Phys. Rev. Lett. \textbf{98}, 108901
(2007).

\noindent \lbrack 23] T. Ivezi\'{c}, Phys. Scr. \textbf{82,} 055007 (2010).

\noindent \lbrack 24] R. D. Blandford and K. S. Thorne, \textit{Applications
of classical physics} (California Institute of Technology, 2002-2003).

\noindent \lbrack 25] D. J. Griffiths, \textit{Introduction to
Electrodynamics} (Pearson, Boston, 2013) 4th ed.

\noindent \lbrack 26] T. Ivezi\'{c}, Phys. Scr.\textit{\ }\textbf{81,}
025001 (2010).

\noindent \lbrack 27] C. Leubner, K. Aufinger and P. Krumm, Eur. J. Physics 
\textbf{13}, 170 (1992).

\noindent \lbrack 28] Z. Oziewicz, J. Phys.: Conf. Ser. \textbf{330}, 012012
(2011).

\noindent \lbrack 29] P. R. Halmos, \textit{Finite-Dimensional Vector Spaces}
( Springer-Verlag, New York Berlin Heidelberg, 1987).

\noindent \lbrack 30] N Bourbaki, \textit{Elements of Mathematics Algebra I }%
(Hermann: Paris; Addison-Wesley: Reading, Massachusetts, 1974).

\noindent \lbrack 31] T. Ivezi\'{c}, \textit{J. Phys.: Conf. Ser}. \textbf{%
845}, 012013 (2017); arXiv: 1101.3292.

\noindent \lbrack 32] T. Ivezi\'{c}, Int. J. Mod. Phys.\textit{\ B} \textbf{%
26}, 1250040 (2012).

\noindent \lbrack 33] T. Ivezi\'{c}, Found. Phys. Lett. \textbf{18,} 401
(2005); arXiv: physics/0412167.

\noindent \lbrack 34] M. Ludvigsen, \textit{General\ Relativity}, \textit{A
Geometric Approach }(Cambridge University Press, Cambridge, 1999).

\noindent \lbrack 35] S. Sonego and M. A. J. Abramowicz, J. Math. Phys. 
\textbf{39}, 3158 (1998).

\noindent \lbrack 36] D. A. T. Vanzella, Phys. Rev. Lett. \textbf{110},
089401 (2013).

\noindent \lbrack 37] H. N. N\'{u}\~{n}ez Y\'{e}pez, A. L. Salas Brito and
C. A. Vargas, Revista Mexicana de F\'{\i}sica \textbf{34}, 636 (1988).

\noindent \lbrack 38] S. Esposito, Found.\ Phys.\ \textbf{28}, 231 (1998).

\noindent \lbrack 39] J. Anandan, Phys. Rev. Lett. \textbf{85}, 1354 (2000).

\noindent \lbrack 40] P. Hillion, Phys. Rev. E \textbf{48,} 3060 (1993).

\noindent \lbrack 41] C. M\o ller, \textit{The Theory of Relativity}, 2nd
ed. (Clarendon Press, Oxford, 1972).

\noindent \lbrack 42] R. M. Wald, \textit{General Relativity} (The
University of Chicago Press, Chicago, 1984).

\noindent \lbrack 43] D. A. T. Vanzella, G. E. A. Matsas and H. W. Crater,
Am. J. Phys. \textbf{64},\textbf{\ }1075 (1996).

\noindent \lbrack 44] F. W. Hehl and Yu. N. Obukhov, \textit{Foundations of
Classical Electrodynamics: Charge, flux, and metric} (Birkh\"{a}user,
Boston, 2003).

\noindent \lbrack 45] T. Ivezi\'{c}, arXiv: hep-th/0207250 (2002); arXiv:
hep-ph/0205277 (2002).

\noindent \lbrack 46] T. Ivezi\'{c}, Found. Phys. Lett. \textbf{12}, 105
(1999).

\noindent \lbrack 47] Kip S. Thorne and Roger D. Blandford, \textit{Modern
Classical Physics: Optics, Fluids, Plasmas, Elasticity, Relativity, and
Statistical Physics} (Princeton University Press, Princeton, 2017).

\noindent \lbrack 48] T. Ivezi\'{c}, Found. Phys. \textbf{36}, 1511 (2006)

\noindent \lbrack 49] Fizika A\textit{\ }\textbf{16,} 207 (2007).

\noindent \lbrack 50] J. D. Jackson, Am. J. Phys. \textbf{72}, 1484 (2004).

\noindent \lbrack 51] O. D. Jefimenko, J. Phys. A: Math. Gen. \textbf{32},
3755 (1999).

\noindent \lbrack 52] M. Mansuripur, Phys. Rev. Lett. \textbf{98}, 193901
(2012)

\noindent \lbrack 53] Proc. SPIE \textbf{8455},\textbf{\ }845512 (2012)

\noindent \lbrack 54] Phys. Rev. Lett. \textbf{110}, 089405 (2013).

\noindent \lbrack 55] T. Ivezi\'{c}, Found. Phys. \textbf{37}, 747 (2007)

\noindent \lbrack 56] T. Ivezi\'{c}, arXiv: 1212.4684 (2012).

\noindent \lbrack 57] D. A. T. Vanzella, Phys. Rev. Lett. \textbf{110},
089401 (2013)

\noindent \lbrack 58] S. M. Barnett, Phys. Rev. Lett. \textbf{110}, 089402
(2013);

\noindent \lbrack 59] P. L. Saldanha, Phys. Rev. Lett. \textbf{110}, 089403
(2013);

\noindent \lbrack 60] M. Khorrami, Phys. Rev. Lett. \textbf{110}, 089404
(2013)

\noindent \lbrack 61] D. J. Griffiths and V. Hnizdo, Am. J. Phys. \textbf{81}%
, 570 (2013).

\noindent \lbrack 62] T. Ivezi\'{c}, arXiv: 1508.04802 (2016).

\noindent \lbrack 63] F. W. Hehl, Ann. Phys. \textbf{17}, 691 (2008).

\noindent \lbrack 64] A. Gamba, Am. J. Phys. \textbf{35}, 83 (1967).

\end{document}